\documentclass{elsart}
\usepackage{graphicx,amssymb,amsfonts,amsmath,hyperref}

\newcommand{\mcm}[3]{\newcommand{#1}[#2]{{\ensuremath{#3}}}}

\usepackage{latexsym}
\usepackage{amssymb}

\mcm{\blank}{0}{(\emptybk)} \mcm{\dashbk}{0}{\mbox{---}}
\mcm{\emptybk}{0}{\:\:} \mcm{\hyph}{0}{\mbox{-}}

\mcm{\diagspace}{0}{\mbox{\hspace{2em}}}

\mcm{\cat}{1}{\mc{#1}} \mcm{\fcat}{1}{\mb{#1}}
\mcm{\mc}{1}{\mathcal{#1}} \mcm{\mr}{1}{\mathrm{#1}}
\mcm{\mi}{1}{\mathit{#1}} \mcm{\mb}{1}{\mathbf{#1}}
\mcm{\scat}{1}{\Bbb{#1}} \mcm{\twid}{1}{\widetilde{#1}}

\mcm{\elt}{0}{\in} \mcm{\sub}{0}{\,\subseteq\,}
\mcm{\such}{0}{\:|\:} \mcm{\without}{0}{\setminus}

\mcm{\atsr}{0}{\Box} \mcm{\eqv}{0}{\,\simeq\,}
\mcm{\iso}{0}{\,\cong\,}
\mcm{\of}{0}{\raisebox{0.2mm}{\ensuremath{\scriptstyle\circ}}}

\mcm{\bdry}{0}{\partial}

\mcm{\Bee}{0}{\cat{B}} \mcm{\Beep}{0}{\cat{B'}}
\mcm{\Eee}{0}{\cat{E}} \mcm{\Eeep}{0}{\cat{E'}}

\mcm{\Ess}{0}{\cat{S}} \mcm{\Tee}{0}{\cat{T}}
\mcm{\Teep}{0}{\cat{T'}} \mcm{\Stee}{0}{\scat{T}}
\mcm{\Steep}{0}{\scat{T'}}

\mcm{\blbk}{0}{\blank^{\blob}}
\mcm{\blob}{0}{\scriptscriptstyle{\bullet}}
\mcm{\stbk}{0}{\blank^{*}} \mcm{\ubl}{0}{{}^{\blob}}
\mcm{\ust}{0}{{}^{*}}

\mcm{\Cartpr}{0}{\pr{\Eee}{T}} \mcm{\Cartprp}{0}{\pr{\Eeep}{T'}}
\mcm{\Mnd}{0}{\triple{T}{\eta}{\mu}}
\mcm{\Zeropr}{0}{\pr{\Set}{\id}}

\mcm{\dopset}{0}{\ftrcat{\Delta^{\op}}{\Set}}
\mcm{\tropset}{0}{\ftrcat{\fcat{TR}^{\op}}{\Set}}

\mcm{\cod}{0}{\mr{cod}} \mcm{\dom}{0}{\mr{dom}}
\mcm{\End}{0}{\mr{End}} \mcm{\Hom}{0}{\mr{Hom}}
\mcm{\ob}{0}{\mr{ob}\,} \mcm{\op}{0}{\mr{op}}

\mcm{\comp}{0}{\mi{comp}} \mcm{\id}{0}{\mi{id}}
\mcm{\ids}{0}{\mi{ids}} \mcm{\mult}{0}{\mi{mult}}
\mcm{\unit}{0}{\mi{unit}}

\mcm{\Ab}{0}{\fcat{Ab}} \mcm{\Alg}{0}{\fcat{Alg}}
\mcm{\Bim}{1}{\fcat{Bim}(#1)} \mcm{\Cat}{0}{\fcat{Cat}}
\mcm{\Cay}{0}{\fcat{Cay}} \mcm{\Cpn}{1}{\pr{\Set/S_{#1}}{T_{#1}}}
\mcm{\fc}{0}{\fcat{fc}} \mcm{\fm}{0}{\fcat{fm}}
\mcm{\Graph}{0}{\fcat{Graph}} \mcm{\Gy}{0}{\fcat{Gy}}
\mcm{\Hpn}{1}{\pr{\Eee_{#1}}{P_{#1}}} \mcm{\Mon}{0}{\mb{Mon}}
\mcm{\Multicat}{0}{\fcat{Multicat}} \mcm{\One}{0}{\fcat{1}}
\mcm{\PD}{1}{\fcat{PD}_{#1}} \mcm{\Prof}{0}{\fcat{Prof}}
\mcm{\Set}{0}{\fcat{Set}} \mcm{\Span}{0}{\fcat{Span}}
\mcm{\Ssq}{0}{\fcat{Ssq}} \mcm{\Struc}{0}{\fcat{Struc}}
\mcm{\Sym}{0}{\fcat{Sym}} \mcm{\TR}{1}{\fcat{TR}(#1)}
\mcm{\Tr}{0}{\fcat{Tr}} \mcm{\Twocat}{0}{\fcat{2\hyph\Cat}}

\mcm{\integers}{0}{\mathbb{Z}}

\mcm{\range}{2}{#1,\,\ldots\,,#2}
\mcm{\bftuple}{2}{\tuplebts{\range{#1}{#2}}}
\mcm{\tuple}{3}{\tuplebts{\range{#1,#2}{#3}}}
\mcm{\rttuple}{1}{\tuplebts{\,\ldots\,,#1}}
\mcm{\abftuple}{2}{\atuplebts{\range{#1}{#2}}}
\mcm{\atuple}{3}{\atuplebts{\range{#1,#2}{#3}}}
\mcm{\arttuple}{1}{\atuplebts{\,\ldots\,,#1}}
\mcm{\sqbftuple}{2}{\obt\range{#1}{#2}\cbt}
\mcm{\pr}{2}{\tuplebts{#1,#2}}
\mcm{\triple}{3}{\tuplebts{#1,#2,#3}}

\mcm{\eend}{2}{#1[#2]} \mcm{\ehom}{3}{#1[#2,#3]}
\mcm{\ftrcat}{2}{[#1,#2]} \mcm{\homset}{3}{#1(#2,#3)}
\mcm{\multihom}{3}{#1(#2;#3)}
\mcm{\relhom}{5}{#1_{#2}(\range{#3}{#4};#5)}

\mcm{\go}{0}{\rTo} \mcm{\goby}{1}{\rTo^{#1}}
\mcm{\goesto}{0}{\,\longmapsto\,} \mcm{\goiso}{0}{\goby{\diso}}
\mcm{\monic}{0}{\rMonic} \mcm{\og}{0}{\lTo}
\mcm{\ogby}{1}{\lTo^{#1}}

\mcm{\gph}{2}{\spn{#1}{T #2}{#2}} \mcm{\graph}{4}{\spaan{#1}{T
#2}{#2}{#3}{#4}} \mcm{\oppair}{2}{\stackrel{\rTo^{#1}}{\lTo_{#2}}}
\mcm{\parpair}{2}{\stackrel{\rTo^{#1}}{\rTo_{#2}}}
\mcm{\spn}{3}{#2 \og #1 \go #3} \mcm{\spaan}{5}{#2 \ogby{#4} #1
\goby{#5} #3}

\mcm{\bktdvslob}{3}
    {\left(
    \begin{diagram}[height=1.5em]
    #1      \\
    \dTo>{\,#2} \\
    #3      \\
    \end{diagram}
    \right)}
\mcm{\slob}{3}{(#1 \goby{#2} #3)} \mcm{\vslob}{3}
    {\left.
    \begin{diagram}[height=1.5em]
    #1      \\
    \dTo>{\,#2} \\
    #3      \\
    \end{diagram}
    \right.}

\newenvironment{tree}
    {\begin{diagram}[height=1em,width=.75em,abut,noPS,tight]}
    {\end{diagram}}

\mcm{\enode}{0}{\circ}

\mcm{\nl}{1}{\stackrel{\textstyle #1}{\node}}
\mcm{\node}{0}{\bullet}

\mcm{\utree}{0}{\node}

\mcm{\diso}{0}{\sim}

\mcm{\vdiso}{0}{\wr}

\mcm{\nat}{0}{\mathbb{N}}

\mcm{\Onepr}{0}{\pr{\Graph}{\fc}}
\newlength{\nllwidth}
\newlength{\nllheight}
\newcommand{\stackbelow}[2]{%
\settowidth{\nllwidth}{\ensuremath{#1}\ensuremath{#2}}%
\settoheight{\nllheight}{\ensuremath{#2}}%
\addtolength{\nllheight}{.3ex}%
\mbox{%
\ensuremath{#1}%
\hspace{-.5\nllwidth}%
\raisebox{-1\nllheight}{\ensuremath{#2}}}}
\mcm{\nlal}{2}{\stackbelow{\nl{#1}}{#2}}
\mcm{\nll}{1}{\stackbelow{\node}{#1}} \mcm{\wun}{0}{\fcat{1}}
\mcm{\atuplebts}{1}{\langle #1 \rangle} \mcm{\tuplebts}{1}{(#1)}
\mcm{\bo}{0}{(} \mcm{\bc}{0}{)}
\mcm{\UBilax}{0}{\fcat{UBicat}_\mr{lax}}
\mcm{\UBiwk}{0}{\fcat{UBicat}_\mr{wk}}
\mcm{\UBistr}{0}{\fcat{UBicat}_\mr{str}}
\mcm{\Bilax}{0}{\fcat{Bicat}_\mr{lax}}
\mcm{\Biwk}{0}{\fcat{Bicat}_\mr{wk}}
\mcm{\Bistr}{0}{\fcat{Bicat}_\mr{str}} \mcm{\rotsub}{0}{\cup
\raisebox{0.1em}{$\scriptstyle{|}$}} \mcm{\pd}{0}{\fcat{pd}}
\mcm{\rep}{1}{\widehat{#1}} \mcm{\ovln}{1}{\overline{#1}}
\mcm{\Gph}{0}{\fcat{Gph}} \mcm{\tr}{0}{\fcat{tr}}

\mcm{\ladj}{0}{\,\dashv\,} \mcm{\zeropd}{0}{\node}
    {\end{diagram}}
\mcm{\END}{0}{\fcat{End}} \mcm{\HOM}{0}{\fcat{Hom}}

\newlength{\gwidth} 
\newlength{\gvert}  
\newlength{\gdrop}  
\newlength{\gbaredrop}  
\newlength{\goffset}    
\newlength{\gtemp}  

\newcommand{\present}[1]{%
\makebox[1\gwidth]{%
\rule[-1\gdrop]{0ex}{1\gvert}%
\raisebox{-1\gbaredrop}{#1}}}

\newcommand{\presentl}[1]{%
\makebox[1\gwidth][l]{%
\rule[-1\gdrop]{0ex}{1\gvert}%
\raisebox{-1\gbaredrop}{#1}}}

\newcommand{\presentr}[1]{%
\makebox[1\gwidth][r]{%
\rule[-1\gdrop]{0ex}{1\gvert}%
\raisebox{-1\gbaredrop}{#1}}}

\newcommand{\ginitdims}[2]{
\setlength{\unitlength}{1em}
\setlength{\goffset}{.25\unitlength}
\setlength{\gwidth}{#1\unitlength}
\setlength{\gvert}{#2\unitlength}
\setlength{\gdrop}{.5\gvert}
\addtolength{\gdrop}{-1\goffset}
\setlength{\gbaredrop}{1\gdrop}
\addtolength{\gvert}{.6\unitlength}
\addtolength{\gdrop}{.3\unitlength}}    

\newcommand{\cinitdims}[2]{
\setlength{\unitlength}{1em}
\setlength{\goffset}{.35\unitlength}
\setlength{\gwidth}{#1\unitlength}
\setlength{\gvert}{#2\unitlength}
\setlength{\gdrop}{.5\gvert}
\addtolength{\gdrop}{-1\goffset}
\setlength{\gbaredrop}{1\gdrop}
\addtolength{\gvert}{.6\unitlength}
\addtolength{\gdrop}{.3\unitlength}}    

\newcommand{\gsinitdims}[2]{
\setlength{\unitlength}{0.5em}
\setlength{\goffset}{.25\unitlength}
\setlength{\gwidth}{#1\unitlength}
\setlength{\gvert}{#2\unitlength}
\setlength{\gdrop}{.5\gvert}
\addtolength{\gdrop}{-1\goffset}
\setlength{\gbaredrop}{1\gdrop}
\addtolength{\gvert}{.6\unitlength}
\addtolength{\gdrop}{.3\unitlength}}    

\newcommand{\sidespic}[1]{%
\settowidth{\gtemp}{\ensuremath{#1}}%
\addtolength{\gwidth}{1\gtemp}}

\newcommand{\abovepic}[1]{%
\settoheight{\gtemp}{\ensuremath{#1}}%
\addtolength{\gvert}{1\gtemp}%
\settodepth{\gtemp}{\ensuremath{#1}}%
\addtolength{\gvert}{1\gtemp}}

\newcommand{\belowpic}[1]{%
\settoheight{\gtemp}{\ensuremath{#1}}%
\addtolength{\gvert}{1\gtemp}%
\addtolength{\gdrop}{1\gtemp}%
\settodepth{\gtemp}{\ensuremath{#1}}%
\addtolength{\gvert}{1\gtemp}%
\addtolength{\gdrop}{1\gtemp}}

\newcommand{\cell}[4]{\put(#1,#2){\makebox(0,0)[#3]{\ensuremath{#4}}}}
\mcm{\zmark}{0}{\scriptstyle{\bullet}}

\newcommand{\pregfst}[1]{%
\begin{picture}(0.5,0.2)(-0.5,-0.2)%
\cell{-0.1}{-0.2}{tr}{#1}%
\cell{0}{0}{c}{\zmark}%
\end{picture}}

\mcm{\gfst}{1}{%
\ginitdims{0.5}{0.4}%
\sidespic{#1}%
\belowpic{#1}%
\presentr{\pregfst{#1}}}

\newcommand{\preglst}[1]{%
\begin{picture}(0.5,0.2)(0,-0.2)%
\cell{0.1}{-0.2}{tl}{#1}%
\cell{0.05}{0}{c}{\zmark}%
\end{picture}}

\mcm{\glst}{1}{%
\ginitdims{.5}{.4}%
\sidespic{#1}%
\belowpic{#1}%
\presentl{\preglst{#1}}}

\newcommand{\preglft}[1]{%
\begin{picture}(0,0.2)(0,-0.2)%
\cell{-0.1}{-0.2}{tr}{#1}%
\cell{0.05}{0}{c}{\zmark}%
\end{picture}}

\mcm{\glft}{1}{%
\ginitdims{0}{.4}%
\belowpic{#1}%
\present{\preglft{#1}}}

\newcommand{\pregrgt}[1]{%
\begin{picture}(0,0.2)(0,-0.2)%
\cell{0.1}{-0.2}{tl}{#1}%
\cell{0.05}{0}{c}{\zmark}%
\end{picture}}

\mcm{\grgt}{1}{%
\ginitdims{0}{.4}%
\belowpic{#1}%
\present{\pregrgt{#1}}}

\newcommand{\pregblw}[1]{%
\begin{picture}(0,0.3)(0,-0.3)
\cell{0}{-0.3}{t}{#1}%
\cell{0.05}{0}{c}{\zmark}%
\end{picture}}

\mcm{\gblw}{1}{%
\ginitdims{0}{.6}%
\belowpic{#1}%
\present{\pregblw{#1}}}

\newcommand{\pregfbw}[1]{%
\begin{picture}(0,0.65)(0,-0.65)
\cell{0}{-0.65}{t}{#1}%
\cell{0.05}{0}{c}{\zmark}%
\end{picture}}

\mcm{\gfbw}{1}{%
\ginitdims{0}{1.3}%
\belowpic{#1}%
\present{\pregfbw{#1}}}

\newcommand{\pregzero}[1]{%
\begin{picture}(0.8,0.4)(-0.4,-0.4)
\cell{0}{-0.4}{t}{#1}%
\cell{0}{0}{c}{\zmark}%
\end{picture}}

\mcm{\gzero}{1}{%
\ginitdims{0.8}{.6}%
\belowpic{#1}%
\sidespic{#1}%
\present{\pregzero{#1}}}

\newcommand{\pregone}[1]{%
\begin{picture}(5,0.4)(0,-0.2)%
\cell{2.5}{0.2}{b}{#1}%
\put(0,0){\vector(1,0){5}}%
\end{picture}}

\mcm{\gone}{1}{%
\ginitdims{5}{0.4}%
\abovepic{#1}%
\present{\pregone{#1}}}

\newcommand{\pregtwo}[3]{%
\begin{picture}(5,3.4)(0,-0.2)%
\cell{2.5}{3.2}{b}{#1}%
\cell{2.5}{-.2}{t}{#2}%
\cell{2.7}{1.5}{l}{#3}%
\qbezier(0,1.5)(2.5,4.5)(5,1.5)%
\qbezier(0,1.5)(2.5,-1.5)(5,1.5)%
\put(5,1.5){\vector(1,-1){0}}%
\put(5,1.5){\vector(1,1){0}}%
\put(2.5,2.5){\vector(0,-1){2}}%
\end{picture}}

\mcm{\gtwo}{3}{%
\ginitdims{5}{3.4}%
\abovepic{#1}%
\belowpic{#2}%
\present{\pregtwo{#1}{#2}{#3}}}

\newcommand{\pregthree}[5]{%
\begin{picture}(5,5.4)(0,-1.2)%
\cell{2.5}{4.2}{b}{#1}%
\cell{1.5}{1.7}{b}{#2}%
\cell{2.5}{-1.2}{t}{#3}%
\cell{2.7}{2.75}{l}{#4}%
\cell{2.7}{0.25}{l}{#5}%
\qbezier(0,1.5)(2.5,6.5)(5,1.5)%
\qbezier(0,1.5)(2.5,-3.5)(5,1.5)%
\put(0,1.5){\vector(1,0){5}}%
\put(2.5,3.5){\vector(0,-1){1.5}}%
\put(2.5,1){\vector(0,-1){1.5}}%
\put(5,1.5){\vector(1,-3){0}}%
\put(5,1.5){\vector(1,3){0}}%
\end{picture}}

\mcm{\gthree}{5}{%
\ginitdims{5}{5.4}%
\abovepic{#1}%
\belowpic{#3}%
\present{\pregthree{#1}{#2}{#3}{#4}{#5}}}

\newcommand{\pregfour}[7]{%
\begin{picture}(5,8.4)(0,-2.7)%
\cell{2.5}{5.7}{b}{#1}%
\cell{1.5}{2.8}{b}{#2}%
\cell{1.5}{0.2}{t}{#3}%
\cell{2.5}{-2.7}{t}{#4}%
\cell{2.7}{4.25}{l}{#5}%
\cell{2.7}{1.5}{l}{#6}%
\cell{2.7}{-1.25}{l}{#7}%
\qbezier(0,1.5)(2.5,9.5)(5,1.5)%
\qbezier(0,1.5)(2.5,4)(5,1.5)%
\qbezier(0,1.5)(2.5,-1)(5,1.5)%
\qbezier(0,1.5)(2.5,-6.5)(5,1.5)%
\put(2.5,5.25){\vector(0,-1){2}}%
\put(2.5,2.5){\vector(0,-1){2}}%
\put(2.5,-0.25){\vector(0,-1){2}}%
\put(5,1.5){\vector(1,-4){0}}%
\put(5,1.5){\vector(4,-3){0}}%
\put(5,1.5){\vector(4,3){0}}%
\put(5,1.5){\vector(1,4){0}}%
\end{picture}}

\mcm{\gfour}{7}{%
\ginitdims{5}{8.4}%
\abovepic{#1}%
\belowpic{#4}%
\present{\pregfour{#1}{#2}{#3}{#4}{#5}{#6}{#7}}}

\newcommand{\pregthreecell}[5]{%
\begin{picture}(8,5)(-4,-2.5)%
\cell{0}{2.5}{b}{#1}%
\cell{0}{-2.5}{t}{#2}%
\cell{-1.7}{0}{r}{#3}%
\cell{1.7}{0}{l}{#4}%
\cell{0}{0.2}{b}{#5}%
\qbezier(-4,0)(0,4.2)(4,0)%
\qbezier(-4,0)(0,-4.2)(4,0)%
\qbezier(-0.5,1.8)(-2.5,0)(-0.5,-1.8)%
\qbezier(0.5,1.8)(2.5,0)(0.5,-1.8)%
\put(-1,0){\vector(1,0){2}}%
\put(4,0){\vector(1,-1){0}}%
\put(4,0){\vector(1,1){0}}%
\put(-0.5,-1.8){\vector(1,-1){0}}%
\put(0.5,-1.8){\vector(-1,-1){0}}%
\end{picture}}

\mcm{\gthreecell}{5}{%
\ginitdims{8}{5}%
\abovepic{#1}%
\belowpic{#2}%
\present{\pregthreecell{#1}{#2}{#3}{#4}{#5}}}

\newcommand{\pregthreecellu}{%
\begin{picture}(5,3.4)(-0.5,-0.2)%
\qbezier(-.5,1.5)(2,4.5)(4.5,1.5)%
\qbezier(-.5,1.5)(2,-1.5)(4.5,1.5)%
\qbezier(1.5,2.7)(0.5,1.5)(1.5,0.3)%
\qbezier(2.5,2.7)(3.5,1.5)(2.5,0.3)%
\put(1.3,1.5){\vector(1,0){1.4}}%
\put(4.5,1.5){\vector(1,-1){0}}%
\put(4.5,1.5){\vector(1,1){0}}%
\put(1.5,0.3){\vector(2,-3){0}}%
\put(2.5,0.3){\vector(-2,-3){0}}%
\end{picture}}

\mcm{\gthreecellu}{0}{%
\ginitdims{5}{3.4}%
\present{\pregthreecellu}}

\newcommand{\pregtwocentre}[3]{%
\begin{picture}(5,3.4)(0,-0.2)%
\cell{2.5}{3.2}{b}{#1}%
\cell{2.5}{-.2}{t}{#2}%
\cell{2.5}{1.5}{c}{#3}%
\qbezier(0,1.5)(2.5,4.5)(5,1.5)%
\qbezier(0,1.5)(2.5,-1.5)(5,1.5)%
\put(5,1.5){\vector(1,-1){0}}%
\put(5,1.5){\vector(1,1){0}}%
\put(2.5,2.5){\vector(0,-1){2}}%
\end{picture}}

\mcm{\gtwocentre}{3}{%
\ginitdims{5}{3.4}%
\abovepic{#1}%
\belowpic{#2}%
\present{\pregtwocentre{#1}{#2}{#3}}}

\newcommand{\pregspecialone}[9]{%
\begin{picture}(8,8)(-4,-4)%
\cell{0}{3.9}{b}{#1}%
\cell{-2}{-0.2}{t}{#2}%
\cell{0}{-3.9}{t}{#3}%
\cell{-1.5}{1.1}{r}{#4}%
\cell{0.2}{1.5}{l}{#5}%
\cell{1.5}{1.1}{l}{#6}%
\cell{0.2}{-2}{l}{#7}%
\cell{-0.9}{2.3}{b}{#8}%
\cell{0.9}{2.3}{b}{#9}%
\qbezier(-4,0)(0,8)(4,0)%
\qbezier(-4,0)(0,-8)(4,0)%
\qbezier(-0.5,3.4)(-3.5,2)(-0.5,0.6)%
\qbezier(0.5,3.4)(3.5,2)(0.5,0.6)%
\put(-4,0){\vector(1,0){8}}%
\put(0,3.4){\vector(0,-1){2.8}}%
\put(0,-0.8){\vector(0,-1){2.4}}%
\put(-1.5,2.2){\vector(1,0){1.2}}%
\put(0.3,2.2){\vector(1,0){1.2}}%
\put(4,0){\vector(1,-2){0}}%
\put(4,0){\vector(1,2){0}}%
\put(-0.5,0.6){\vector(2,-1){0}}%
\put(0.5,0.6){\vector(-2,-1){0}}%
\end{picture}}

\mcm{\gspecialone}{9}{%
\ginitdims{8}{8}%
\abovepic{#1}%
\belowpic{#3}%
\present{\pregspecialone{#1}{#2}{#3}{#4}{#5}{#6}{#7}{#8}{#9}}}

\newcommand{\pregspecialtwo}{%
\begin{picture}(5,3.4)(0,-0.2)%
\qbezier(0,1.5)(2.5,4.5)(5,1.5)%
\qbezier(0,1.5)(2.5,-1.5)(5,1.5)%
\qbezier(1.7,2.5)(0,1.5)(1.7,0.5)%
\qbezier(3.3,2.5)(5,1.5)(3.3,0.5)%
\put(5,1.5){\vector(1,-1){0}}%
\put(5,1.5){\vector(1,1){0}}%
\put(1.7,0.5){\vector(3,-2){0}}%
\put(3.3,0.5){\vector(-3,-2){0}}%
\put(2.5,2.5){\vector(0,-1){2}}%
\put(1.2,1.5){\vector(1,0){1}}%
\put(2.8,1.5){\vector(1,0){1}}%
\end{picture}}

\mcm{\gspecialtwo}{0}{%
\ginitdims{5}{3.4}%
\present{\pregspecialtwo}}

\newcommand{\pregspecialthree}{%
\begin{picture}(5,5.4)(0,-1.2)%
\qbezier(0,1.5)(2.5,6.5)(5,1.5)%
\qbezier(0,1.5)(2.5,-3.5)(5,1.5)%
\qbezier(2,3.5)(1,2.75)(2,2)%
\qbezier(3,3.5)(4,2.75)(3,2)%
\qbezier(2,1)(1,0.25)(2,-0.5)%
\qbezier(3,1)(4,0.25)(3,-0.5)%
\put(0,1.5){\vector(1,0){5}}%
\put(1.5,2.75){\vector(1,0){2}}%
\put(1.5,0.25){\vector(1,0){2}}%
\put(5,1.5){\vector(1,-3){0}}%
\put(5,1.5){\vector(1,3){0}}%
\put(2,2){\vector(1,-1){0}}%
\put(3,2){\vector(-1,-1){0}}%
\put(2,-0.5){\vector(1,-1){0}}%
\put(3,-0.5){\vector(-1,-1){0}}%
\end{picture}}

\mcm{\gspecialthree}{0}{%
\ginitdims{5}{5.4}%
\present{\pregspecialthree}}

\newcommand{\pregonew}[1]{%
\begin{picture}(8,0.4)(0,-0.2)%
\cell{4}{0.2}{b}{#1}%
\put(0,0){\vector(1,0){8}}%
\end{picture}}

\mcm{\gonew}{1}{%
\ginitdims{8}{0.4}%
\abovepic{#1}%
\present{\pregonew{#1}}}

\mcm{\gzersu}{0}{%
\gsinitdims{0}{.6}%
\present{\pregblw{}}}

\mcm{\gonesu}{0}{%
\gsinitdims{5}{0.4}%
\present{\pregone{}}}

\mcm{\gtwosu}{0}{%
\gsinitdims{5}{3.4}%
\present{\pregtwo{}{}{}}}

\mcm{\gthreesu}{0}{%
\gsinitdims{5}{5.4}%
\present{\pregthree{}{}{}{}{}}}

\mcm{\gfoursu}{0}{%
\gsinitdims{5}{8.4}%
\present{\pregfour{}{}{}{}{}{}{}}}

\newcommand{\precone}[1]{%
\begin{picture}(4.2,0.4)(-0.3,-0.2)%
\cell{1.8}{0.2}{b}{#1}%
\put(0,0){\vector(1,0){3.6}}%
\end{picture}}

\mcm{\cone}{1}{%
\cinitdims{4.2}{0.4}%
\abovepic{#1}%
\present{\precone{#1}}}

\mcm{\gfstsu}{0}{%
\gsinitdims{0.5}{0.4}%
\presentr{\pregfst{}}}

\mcm{\glstsu}{0}{%
\gsinitdims{0.5}{0.4}%
\presentl{\preglst{}}}

\newcommand{\prectwodbl}[3]%
{\begin{picture}(4.2,3.4)(-0.1,-0.2)%
\cell{2}{3.2}{b}{#1}%
\cell{2}{-0.2}{t}{#2}%
\cell{2.3}{1.5}{l}{#3}%
\qbezier(0,2)(2,4)(4,2)%
\qbezier(0,1)(2,-1)(4,1)%
\put(4,2){\vector(1,-1){0}}%
\put(4,1){\vector(1,1){0}}%
\put(1.9,2.5){\line(0,-1){1.8}}%
\put(2.1,2.5){\line(0,-1){1.8}}%
\cell{2.01}{0.4}{b}{\vee}%
\end{picture}}

\mcm{\ctwodbl}{3}{%
\cinitdims{4.2}{3.4}%
\abovepic{#1}%
\belowpic{#2}%
\present{\prectwodbl{#1}{#2}{#3}}}

\newcommand{\precthreedbl}[5]{%
\begin{picture}(4.2,5.4)(-0.1,-0.2)%
\cell{2}{5.2}{b}{#1}%
\cell{1}{2.7}{b}{#2}%
\cell{2}{-.2}{t}{#3}%
\cell{2.3}{3.75}{l}{#4}%
\cell{2.3}{1.25}{l}{#5}%
\qbezier(0,3)(2,7)(4,3)%
\qbezier(0,2)(2,-2)(4,2)%
\put(0,2.5){\vector(1,0){4}}%
\put(1.9,4.5){\line(0,-1){1.3}}%
\put(2.1,4.5){\line(0,-1){1.3}}%
\cell{2.01}{2.9}{b}{\vee}%
\put(1.9,2){\line(0,-1){1.3}}%
\put(2.1,2){\line(0,-1){1.3}}%
\cell{2.01}{0.4}{b}{\vee}%
\put(4,3){\vector(1,-3){0}}%
\put(4,2){\vector(1,3){0}}%
\end{picture}}

\mcm{\cthreedbl}{5}{%
\cinitdims{4.2}{5.4}%
\abovepic{#1}%
\belowpic{#3}%
\present{\precthreedbl{#1}{#2}{#3}{#4}{#5}}}

\newcommand{\precthreecelltrp}[5]{%
\begin{picture}(8.2,5)(-4.1,-2.5)%
\cell{0}{2.5}{b}{#1}%
\cell{0}{-2.5}{t}{#2}%
\cell{-1.8}{0}{r}{#3}%
\cell{1.8}{0}{l}{#4}%
\cell{0}{0.3}{b}{#5}%
\qbezier(-4,0.5)(0,4)(4,0.5)%
\qbezier(-4,-0.5)(0,-4)(4,-0.5)%
\qbezier(-0.6,2)(-2.6,0)(-0.6,-2)%
\qbezier(-0.4,2)(-2.4,0)(-0.5,-1.9)%
\cell{-0.6}{-2}{b}{\lrcorner}%
\qbezier(0.4,2)(2.4,0)(0.5,-1.9)%
\qbezier(0.6,2)(2.6,0)(0.6,-2)%
\cell{0.65}{-2}{b}{\llcorner}%
\put(-1,0.15){\line(1,0){1.7}}%
\put(-1,0){\line(1,0){2}}%
\put(-1,-0.15){\line(1,0){1.7}}%
\cell{1.15}{0}{r}{>}%
\put(4,0.5){\vector(1,-1){0}}%
\put(4,-0.5){\vector(1,1){0}}%
\end{picture}}

\mcm{\cthreecelltrp}{5}{%
\cinitdims{8.2}{5}%
\abovepic{#1}%
\belowpic{#2}%
\present{\precthreecelltrp{#1}{#2}{#3}{#4}{#5}}}

\newcommand{\prectwo}[3]%
{\begin{picture}(4.2,3.4)(-0.1,-0.2)%
\cell{2}{3.2}{b}{#1}%
\cell{2}{-0.2}{t}{#2}%
\cell{2.2}{1.5}{l}{#3}%
\qbezier(0,2)(2,4)(4,2)%
\qbezier(0,1)(2,-1)(4,1)%
\put(4,2){\vector(1,-1){0}}%
\put(4,1){\vector(1,1){0}}%
\put(2,2.5){\vector(0,-1){2}}%
\end{picture}}

\mcm{\ctwo}{3}{%
\cinitdims{4.2}{3.4}%
\abovepic{#1}%
\belowpic{#2}%
\present{\prectwo{#1}{#2}{#3}}}

\newcommand{\precthree}[5]{%
\begin{picture}(4.2,5.4)(-0.1,-0.2)%
\cell{2}{5.2}{b}{#1}%
\cell{1}{2.7}{b}{#2}%
\cell{2}{-.2}{t}{#3}%
\cell{2.2}{3.75}{l}{#4}%
\cell{2.2}{1.25}{l}{#5}%
\qbezier(0,3)(2,7)(4,3)%
\qbezier(0,2)(2,-2)(4,2)%
\put(0,2.5){\vector(1,0){4}}%
\put(2,4.5){\vector(0,-1){1.5}}%
\put(2,2){\vector(0,-1){1.5}}%
\put(4,3){\vector(1,-3){0}}%
\put(4,2){\vector(1,3){0}}%
\end{picture}}

\mcm{\cthree}{5}{%
\cinitdims{4.2}{5.4}%
\abovepic{#1}%
\belowpic{#3}%
\present{\precthree{#1}{#2}{#3}{#4}{#5}}}

\newcommand{\prectwoop}[3]%
{\begin{picture}(4.2,3.4)(-0.1,-0.2)%
\cell{2}{3.2}{b}{#1}%
\cell{2}{-0.2}{t}{#2}%
\cell{2.2}{1.5}{l}{#3}%
\qbezier(0,2)(2,4)(4,2)%
\qbezier(0,1)(2,-1)(4,1)%
\put(0,2){\vector(-1,-1){0}}%
\put(0,1){\vector(-1,1){0}}%
\put(2,2.5){\vector(0,-1){2}}%
\end{picture}}

\mcm{\ctwoop}{3}{%
\cinitdims{4.2}{3.4}%
\abovepic{#1}%
\belowpic{#2}%
\present{\prectwoop{#1}{#2}{#3}}}

\newcommand{\prectwopar}[4]{%
\begin{picture}(4.2,3.4)(-0.1,-0.2)%
\cell{2}{3.2}{b}{#1}%
\cell{2}{-0.2}{t}{#2}%
\cell{1.6}{1.5}{r}{#3}%
\cell{2.4}{1.5}{l}{#4}%
\qbezier(0,2)(2,4)(4,2)%
\qbezier(0,1)(2,-1)(4,1)%
\put(4,2){\vector(1,-1){0}}%
\put(4,1){\vector(1,1){0}}%
\put(1.8,2.5){\vector(0,-1){2}}%
\put(2.2,2.5){\vector(0,-1){2}}%
\end{picture}}

\mcm{\ctwopar}{4}{%
\cinitdims{4.2}{3.4}%
\abovepic{#1}%
\belowpic{#2}%
\present{\prectwopar{#1}{#2}{#3}{#4}}}

\newcommand{\precthreein}[5]{%
\begin{picture}(4.2,5.4)(-0.1,-0.2)%
\cell{2}{5.2}{b}{#1}%
\cell{1}{2.7}{b}{#2}%
\cell{2}{-.2}{t}{#3}%
\cell{2.2}{3.75}{l}{#4}%
\cell{2.2}{1.25}{l}{#5}%
\qbezier(0,3)(2,7)(4,3)%
\qbezier(0,2)(2,-2)(4,2)%
\put(0,2.5){\vector(1,0){4}}%
\put(2,4.5){\vector(0,-1){1.5}}%
\put(2,0.5){\vector(0,1){1.5}}%
\put(4,3){\vector(1,-3){0}}%
\put(4,2){\vector(1,3){0}}%
\end{picture}}

\mcm{\cthreein}{5}{%
\cinitdims{4.2}{5.4}%
\abovepic{#1}%
\belowpic{#3}%
\present{\precthreein{#1}{#2}{#3}{#4}{#5}}}

\newcommand{\precthreecell}[5]{%
\begin{picture}(8.2,5)(-4.1,-2.5)%
\cell{0}{2.5}{b}{#1}%
\cell{0}{-2.5}{t}{#2}%
\cell{-1.7}{0}{r}{#3}%
\cell{1.7}{0}{l}{#4}%
\cell{0}{0.2}{b}{#5}%
\qbezier(-4,0.5)(0,4)(4,0.5)%
\qbezier(-4,-0.5)(0,-4)(4,-0.5)%
\qbezier(-0.5,2)(-2.5,0)(-0.5,-2)%
\qbezier(0.5,2)(2.5,0)(0.5,-2)%
\put(-1,0){\vector(1,0){2}}%
\put(4,0.5){\vector(1,-1){0}}%
\put(4,-0.5){\vector(1,1){0}}%
\put(-0.5,-2){\vector(1,-1){0}}%
\put(0.5,-2){\vector(-1,-1){0}}%
\end{picture}}

\mcm{\cthreecell}{5}{%
\cinitdims{8.2}{5}%
\abovepic{#1}%
\belowpic{#2}%
\present{\precthreecell{#1}{#2}{#3}{#4}{#5}}}

\newcommand{\precthreecellpar}[6]{%
\begin{picture}(8.2,5)(-4.1,-2.5)%
\cell{0}{2.5}{b}{#1}%
\cell{0}{-2.5}{t}{#2}%
\cell{-1.7}{0}{r}{#3}%
\cell{1.7}{0}{l}{#4}%
\cell{0}{0.4}{b}{#5}%
\cell{0}{-0.4}{t}{#6}%
\qbezier(-4,0.5)(0,4)(4,0.5)%
\qbezier(-4,-0.5)(0,-4)(4,-0.5)%
\qbezier(-0.5,2)(-2.5,0)(-0.5,-2)%
\qbezier(0.5,2)(2.5,0)(0.5,-2)%
\put(-1,0.2){\vector(1,0){2}}%
\put(-1,-0.2){\vector(1,0){2}}%
\put(4,0.5){\vector(1,-1){0}}%
\put(4,-0.5){\vector(1,1){0}}%
\put(-0.5,-2){\vector(1,-1){0}}%
\put(0.5,-2){\vector(-1,-1){0}}%
\end{picture}}

\mcm{\cthreecellpar}{6}{%
\cinitdims{8.2}{5}%
\abovepic{#1}%
\belowpic{#2}%
\present{\precthreecellpar{#1}{#2}{#3}{#4}{#5}{#6}}}

\newcommand{\prectwov}[5]{%
\begin{picture}(3.4,4.2)(0.8,0.9)%
\cell{2.5}{5.1}{b}{#1}%
\cell{2.5}{0.9}{t}{#2}%
\cell{0.8}{3}{r}{#3}%
\cell{4.2}{3}{l}{#4}%
\cell{2.5}{3.2}{b}{#5}%
\qbezier(2,5)(0,3)(2,1)%
\qbezier(3,5)(5,3)(3,1)%
\put(2,1){\vector(1,-1){0}}%
\put(3,1){\vector(-1,-1){0}}%
\put(1.5,3){\vector(1,0){2}}%
\end{picture}}

\mcm{\ctwov}{5}{%
\cinitdims{3.4}{4.2}%
\abovepic{#1}%
\belowpic{#2}%
\sidespic{#3}%
\sidespic{#4}%
\present{\prectwov{#1}{#2}{#3}{#4}{#5}}}

\newcommand{\precthreecellv}[7]{%
\begin{picture}(5,8.2)(0.5,-1.6)%
\cell{3}{6.6}{b}{#1}%
\cell{3}{-1.6}{t}{#2}%
\cell{0.5}{2.5}{r}{#3}%
\cell{5.5}{2.5}{l}{#4}%
\cell{3}{4.2}{b}{#5}%
\cell{3}{0.8}{t}{#6}%
\cell{3.2}{2.5}{l}{#7}%
\qbezier(3.5,6.5)(7,2.5)(3.5,-1.5)%
\qbezier(2.5,6.5)(-1,2.5)(2.5,-1.5)%
\put(2.5,-1.5){\vector(1,-1){0}}%
\put(3.5,-1.5){\vector(-1,-1){0}}%
\qbezier(1,3)(3,5)(5,3)%
\qbezier(1,2)(3,0)(5,2)%
\put(5,3){\vector(1,-1){0}}%
\put(5,2){\vector(1,1){0}}%
\put(3,3.5){\vector(0,-1){2}}%
\end{picture}}

\mcm{\cthreecellv}{7}{%
\cinitdims{5}{8.2}%
\abovepic{#1}%
\belowpic{#2}%
\sidespic{#3}%
\sidespic{#4}%
\present{\precthreecellv{#1}{#2}{#3}{#4}{#5}{#6}{#7}}}

\newcommand{\pretopez}[2]{%
\begin{picture}(2.6,2.3)(-1.3,-2.2)%
\cell{0}{-2.2}{t}{#1}%
\cell{0}{-1.2}{c}{#2}%
\qbezier(0,0)(-2,-2)(0,-2)%
\qbezier(0,0)(2,-2)(0,-2)%
\put(0,0){\vector(-1,1){0}}%
\end{picture}}

\mcm{\topez}{2}{%
\ginitdims{2.6}{2.3}%
\belowpic{#1}%
\present{\pretopez{#1}{#2}}}

\newcommand{\pretopea}[3]{%
\begin{picture}(4,1.9)(-2,-0,2)%
\cell{0}{1.7}{b}{#1}%
\cell{0}{-0.2}{t}{#2}%
\cell{0}{0.7}{c}{#3}%
\qbezier(-2,0)(0,3)(2,0)%
\put(-2,0){\vector(1,0){4}}%
\put(2,0){\vector(2,-3){0}}%
\end{picture}}

\mcm{\topea}{3}{%
\ginitdims{4}{1.9}%
\abovepic{#1}%
\belowpic{#2}%
\present{\pretopea{#1}{#2}{#3}}}

\newcommand{\pretopeb}[4]{%
\begin{picture}(4,2.2)(-2,-0.2)%
\cell{-1.1}{1}{br}{#1}%
\cell{1.1}{1}{bl}{#2}%
\cell{0}{-0.2}{t}{#3}%
\cell{0}{0.8}{c}{#4}%
\put(-2,0){\vector(1,1){2}}%
\put(0,2){\vector(1,-1){2}}%
\put(-2,0){\vector(1,0){4}}%
\end{picture}}

\mcm{\topeb}{4}{%
\ginitdims{4}{2.2}%
\belowpic{#3}%
\present{\pretopeb{#1}{#2}{#3}{#4}}}

\newcommand{\pretopec}[5]{%
\begin{picture}(4,2.2)(-2,-0.2)%
\cell{-1.8}{1}{br}{#1}%
\cell{0}{2.2}{b}{#2}%
\cell{1.8}{1}{bl}{#3}%
\cell{0}{-0.2}{t}{#4}%
\cell{0}{0.8}{c}{#5}%
\put(-2,0){\vector(1,2){1}}%
\put(-1,2){\vector(1,0){2}}%
\put(1,2){\vector(1,-2){1}}%
\put(-2,0){\vector(1,0){4}}%
\end{picture}}

\mcm{\topec}{5}{%
\ginitdims{4}{2.2}%
\sidespic{#1}%
\abovepic{#2}%
\sidespic{#3}%
\belowpic{#4}%
\present{\pretopec{#1}{#2}{#3}{#4}{#5}}}

\newcommand{\pretoped}[6]{%
\begin{picture}(4,2.5)(-2,-0.2)%
\cell{-2}{0.6}{br}{#1}%
\cell{-0.7}{2.2}{br}{#2}%
\cell{0.7}{2.2}{bl}{#3}%
\cell{2}{0.6}{bl}{#4}%
\cell{0}{-0.2}{t}{#5}%
\cell{0}{0.8}{c}{#6}%
\put(-2,0){\vector(1,3){0.5}}%
\put(-1.5,1.5){\vector(3,2){1.5}}%
\put(0,2.5){\vector(3,-2){1.5}}%
\put(1.5,1.5){\vector(1,-3){0.5}}%
\put(-2,0){\vector(1,0){4}}%
\end{picture}}

\mcm{\toped}{6}{%
\ginitdims{4}{2.5}%
\sidespic{#1}%
\abovepic{#2}%
\abovepic{#3}%
\sidespic{#4}%
\belowpic{#5}%
\present{\pretoped{#1}{#2}{#3}{#4}{#5}{#6}}}

\newcommand{\pretopeq}[5]{%
\begin{picture}(4,2.5)(-2,-0.2)%
\cell{-2}{0.6}{br}{#1}%
\cell{-1}{2.2}{br}{#2}%
\cell{2}{0.6}{bl}{#3}%
\cell{0}{-0.2}{t}{#4}%
\cell{0}{0.8}{c}{#5}%
\put(-2,0){\vector(1,3){0.5}}%
\put(-1.5,1.5){\vector(1,1){1}}%
\cell{0.9}{2.3}{c}{\ddots}
\put(1.5,1.5){\vector(1,-3){0.5}}%
\put(-2,0){\vector(1,0){4}}%
\end{picture}}

\mcm{\topeq}{5}{%
\ginitdims{4}{2.5}%
\sidespic{#1}%
\abovepic{#2}%
\sidespic{#3}%
\belowpic{#4}%
\present{\pretopeq{#1}{#2}{#3}{#4}{#5}}}

\newcommand{\pretopebase}[1]{%
\begin{picture}(4,0.4)(0,-0.2)%
\cell{2}{0.2}{b}{#1}%
\put(0,0){\vector(1,0){4}}%
\end{picture}}

\mcm{\topebase}{1}{%
\ginitdims{4}{0.4}%
\abovepic{#1}%
\present{\pretopebase{#1}}}

\newcommand{\pretopezs}[2]{%
\begin{picture}(2.6,2.3)(-1.3,-2.2)%
\cell{0}{-2.2}{t}{#1}%
\cell{0}{-1.2}{c}{#2}%
\qbezier(0,0)(-2,-2)(0,-2)%
\qbezier(0,0)(2,-2)(0,-2)%
\end{picture}}

\mcm{\topezs}{2}{%
\ginitdims{2.6}{2.3}%
\belowpic{#1}%
\present{\pretopezs{#1}{#2}}}

\newcommand{\pretopeas}[3]{%
\begin{picture}(4,1.9)(-2,-0,2)%
\cell{0}{1.7}{b}{#1}%
\cell{0}{-0.2}{t}{#2}%
\cell{0}{0.7}{c}{#3}%
\qbezier(-2,0)(0,3)(2,0)%
\put(-2,0){\line(1,0){4}}%
\end{picture}}

\mcm{\topeas}{3}{%
\ginitdims{4}{1.9}%
\abovepic{#1}%
\belowpic{#2}%
\present{\pretopeas{#1}{#2}{#3}}}

\newcommand{\pretopebs}[4]{%
\begin{picture}(4,2.2)(-2,-0.2)%
\cell{-1.1}{1}{br}{#1}%
\cell{1.1}{1}{bl}{#2}%
\cell{0}{-0.2}{t}{#3}%
\cell{0}{0.8}{c}{#4}%
\put(-2,0){\line(1,1){2}}%
\put(0,2){\line(1,-1){2}}%
\put(-2,0){\line(1,0){4}}%
\end{picture}}

\mcm{\topebs}{4}{%
\ginitdims{4}{2.2}%
\belowpic{#3}%
\present{\pretopebs{#1}{#2}{#3}{#4}}}

\newcommand{\pretopecs}[5]{%
\begin{picture}(4,2.2)(-2,-0.2)%
\cell{-1.8}{1}{br}{#1}%
\cell{0}{2.2}{b}{#2}%
\cell{1.8}{1}{bl}{#3}%
\cell{0}{-0.2}{t}{#4}%
\cell{0}{0.8}{c}{#5}%
\put(-2,0){\line(1,2){1}}%
\put(-1,2){\line(1,0){2}}%
\put(1,2){\line(1,-2){1}}%
\put(-2,0){\line(1,0){4}}%
\end{picture}}

\mcm{\topecs}{5}{%
\ginitdims{4}{2.2}%
\sidespic{#1}%
\abovepic{#2}%
\sidespic{#3}%
\belowpic{#4}%
\present{\pretopecs{#1}{#2}{#3}{#4}{#5}}}

\newcommand{\pretopeds}[6]{%
\begin{picture}(4,2.5)(-2,-0.2)%
\cell{-2}{0.6}{br}{#1}%
\cell{-0.7}{2.2}{br}{#2}%
\cell{0.7}{2.2}{bl}{#3}%
\cell{2}{0.6}{bl}{#4}%
\cell{0}{-0.2}{t}{#5}%
\cell{0}{0.8}{c}{#6}%
\put(-2,0){\line(1,3){0.5}}%
\put(-1.5,1.5){\line(3,2){1.5}}%
\put(0,2.5){\line(3,-2){1.5}}%
\put(1.5,1.5){\line(1,-3){0.5}}%
\put(-2,0){\line(1,0){4}}%
\end{picture}}

\mcm{\topeds}{6}{%
\ginitdims{4}{2.5}%
\sidespic{#1}%
\abovepic{#2}%
\abovepic{#3}%
\sidespic{#4}%
\belowpic{#5}%
\present{\pretopeds{#1}{#2}{#3}{#4}{#5}{#6}}}

\newcommand{\pretopeqs}[5]{%
\begin{picture}(4,2.5)(-2,-0.2)%
\cell{-2}{0.6}{br}{#1}%
\cell{-1}{2.2}{br}{#2}%
\cell{2}{0.6}{bl}{#3}%
\cell{0}{-0.2}{t}{#4}%
\cell{0}{0.8}{c}{#5}%
\put(-2,0){\line(1,3){0.5}}%
\put(-1.5,1.5){\line(1,1){1}}%
\cell{0.9}{2.3}{c}{\ddots}
\put(1.5,1.5){\line(1,-3){0.5}}%
\put(-2,0){\line(1,0){4}}%
\end{picture}}

\mcm{\topeqs}{5}{%
\ginitdims{4}{2.5}%
\sidespic{#1}%
\abovepic{#2}%
\sidespic{#3}%
\belowpic{#4}%
\present{\pretopeqs{#1}{#2}{#3}{#4}{#5}}}

\newcommand{\pretopebases}[1]{%
\begin{picture}(4,0.4)(0,-0.2)%
\cell{2}{0.2}{b}{#1}%
\put(0,0){\line(1,0){4}}%
\end{picture}}

\mcm{\topebases}{1}{%
\ginitdims{4}{0.4}%
\abovepic{#1}%
\present{\pretopebases{#1}}}

\newcommand{\pregdots}[6]{%
\begin{picture}(5,8.4)(0,-2.7)%
\cell{2.5}{5.7}{b}{#1}%
\cell{1.5}{2.8}{b}{#2}%
\cell{1.5}{0.2}{t}{#3}%
\cell{2.5}{-2.7}{t}{#4}%
\cell{2.7}{4.25}{l}{#5}%
\cell{2.7}{-1.25}{l}{#6}%
\qbezier(0,1.5)(2.5,9.5)(5,1.5)%
\qbezier(0,1.5)(2.5,4)(5,1.5)%
\qbezier(0,1.5)(2.5,-1)(5,1.5)%
\qbezier(0,1.5)(2.5,-6.5)(5,1.5)%
\put(2.5,5.25){\vector(0,-1){2}}%
\put(2.5,-0.25){\vector(0,-1){2}}%
\cell{2.5}{1.7}{c}{\vdots}%
\put(5,1.5){\vector(1,-4){0}}%
\put(5,1.5){\vector(4,-3){0}}%
\put(5,1.5){\vector(4,3){0}}%
\put(5,1.5){\vector(1,4){0}}%
\end{picture}}

\mcm{\gdots}{6}{%
\ginitdims{5}{8.4}%
\abovepic{#1}%
\belowpic{#4}%
\present{\pregdots{#1}{#2}{#3}{#4}{#5}{#6}}}

\newlength{\volt}
\makeatletter

\def\diagram{\m@th\leftwidth=\z@ \rightwidth=\z@ \topheight=\z@
\botheight=\z@ \setbox\@picbox\hbox\bgroup}

\def\enddiagram{\egroup\wd\@picbox\rightwidth\unitlength
\ht\@picbox\topheight\unitlength \dp\@picbox\botheight\unitlength
\hskip\leftwidth\unitlength\box\@picbox}

\def\bfig{\begin{diagram}}
\def\efig{\end{diagram}}
\newcount\wideness \newcount\leftwidth \newcount\rightwidth
\newcount\highness \newcount\topheight \newcount\botheight

\def\ratchet#1#2{\ifnum#1<#2 \global #1=#2 \fi}

\def\putbox(#1,#2)#3{%
\horsize{\wideness}{#3} \divide\wideness by 2 {\advance\wideness
by #1 \ratchet{\rightwidth}{\wideness}} {\advance\wideness by -#1
\ratchet{\leftwidth}{\wideness}} \vertsize{\highness}{#3}
\divide\highness by 2 {\advance\highness by #2
\ratchet{\topheight}{\highness}} {\advance\highness by -#2
\ratchet{\botheight}{\highness}} \put(#1,#2){\makebox(0,0){$#3$}}}

\def\putlbox(#1,#2)#3{%
\horsize{\wideness}{#3} {\advance\wideness by #1
\ratchet{\rightwidth}{\wideness}} {\ratchet{\leftwidth}{-#1}}
\vertsize{\highness}{#3} \divide\highness by 2 {\advance\highness
by #2 \ratchet{\topheight}{\highness}} {\advance\highness by -#2
\ratchet{\botheight}{\highness}}
\put(#1,#2){\makebox(0,0)[l]{$#3$}}}

\def\putrbox(#1,#2)#3{%
\horsize{\wideness}{#3} {\ratchet{\rightwidth}{#1}}
{\advance\wideness by -#1 \ratchet{\leftwidth}{\wideness}}
\vertsize{\highness}{#3} \divide\highness by 2 {\advance\highness
by #2 \ratchet{\topheight}{\highness}} {\advance\highness by -#2
\ratchet{\botheight}{\highness}}
\put(#1,#2){\makebox(0,0)[r]{$#3$}}}

\def\adjust[#1]{} 

\newcount \coefa
\newcount \coefb
\newcount \coefc
\newcount\tempcounta
\newcount\tempcountb
\newcount\tempcountc
\newcount\tempcountd
\newcount\xext
\newcount\yext
\newcount\xoff
\newcount\yoff
\newcount\gap%
\newcount\arrowtypea
\newcount\arrowtypeb
\newcount\arrowtypec
\newcount\arrowtyped
\newcount\arrowtypee
\newcount\height
\newcount\width
\newcount\xpos
\newcount\ypos
\newcount\run
\newcount\rise
\newcount\arrowlength
\newcount\halflength
\newcount\arrowtype
\newdimen\tempdimen
\newdimen\xlen
\newdimen\ylen
\newsavebox{\tempboxa}%
\newsavebox{\tempboxb}%
\newsavebox{\tempboxc}%

\newdimen\w@dth

\def\setw@dth#1#2{\setbox\z@\hbox{\m@th$#1$}\w@dth=\wd\z@
\setbox\@ne\hbox{\m@th$#2$}\ifnum\w@dth<\wd\@ne \w@dth=\wd\@ne \fi
\advance\w@dth by 1.2em}

\def\t@^#1_#2{\allowbreak\def\n@one{#1}\def\n@two{#2}\mathrel
{\setw@dth{#1}{#2} \mathop{\hbox to
\w@dth{\rightarrowfill}}\limits \ifx\n@one\empty\else
^{\box\z@}\fi \ifx\n@two\empty\else _{\box\@ne}\fi}}
\def\t@@^#1{\@ifnextchar_{\t@^{#1}}{\t@^{#1}_{}}}
\def\to{\@ifnextchar^{\t@@}{\t@@^{}}}

\def\t@left^#1_#2{\def\n@one{#1}\def\n@two{#2}\mathrel{\setw@dth{#1}{#2}
\mathop{\hbox to \w@dth{\leftarrowfill}}\limits
\ifx\n@one\empty\else ^{\box\z@}\fi \ifx\n@two\empty\else
_{\box\@ne}\fi}}
\def\t@@left^#1{\@ifnextchar_{\t@left^{#1}}{\t@left^{#1}_{}}}
\def\toleft{\@ifnextchar^{\t@@left}{\t@@left^{}}}

\def\two@^#1_#2{\allowbreak
\def\n@one{#1}\def\n@two{#2}\mathrel{\setw@dth{#1}{#2}
\mathop{\vcenter{\lineskip\z@\baselineskip\z@
                 \hbox to \w@dth{\rightarrowfill}%
                 \hbox to \w@dth{\rightarrowfill}}%
       }\limits
\ifx\n@one\empty\else ^{\box\z@}\fi \ifx\n@two\empty\else
_{\box\@ne}\fi}}
\def\tw@@^#1{\@ifnextchar _{\two@^{#1}}{\two@^{#1}_{}}}
\def\two{\@ifnextchar ^{\tw@@}{\tw@@^{}}}

\def\tofr@^#1_#2{\def\n@one{#1}\def\n@two{#2}\mathrel{\setw@dth{#1}{#2}
\mathop{\vcenter{\hbox to \w@dth{\rightarrowfill}\kern-1.7ex
                 \hbox to \w@dth{\leftarrowfill}}%
       }\limits
\ifx\n@one\empty\else ^{\box\z@}\fi \ifx\n@two\empty\else
_{\box\@ne}\fi}}
\def\t@fr@^#1{\@ifnextchar_ {\tofr@^{#1}}{\tofr@^{#1}_{}}}
\def\tofro{\@ifnextchar^ {\t@fr@}{\t@fr@^{}}}

\def\mon{\mathop{\m@th\hbox to
      14.6\P@{\lasyb\char'51\hskip-2.1\P@$\arrext$\hss
$\mathord\rightarrow$}}\limits} 
\def\leftmono{\mathrel{\m@th\hbox to
14.6\P@{$\mathord\leftarrow$\hss$\arrext$\hskip-2.1\P@\lasyb\char'50%
}}\limits} 
\mathchardef\arrext="0200       

\def\settypes(#1,#2,#3){\arrowtypea#1 \arrowtypeb#2 \arrowtypec#3}
\def\settoheight#1#2{\setbox\@tempboxa\hbox{#2}#1\ht\@tempboxa\relax}%
\def\settodepth#1#2{\setbox\@tempboxa\hbox{#2}#1\dp\@tempboxa\relax}%
\def\settokens`#1`#2`#3`#4`{%
     \def\tokena{#1}\def\tokenb{#2}\def\tokenc{#3}\def\tokend{#4}}
\def\setsqparms[#1`#2`#3`#4;#5`#6]{%
\arrowtypea #1 \arrowtypeb #2 \arrowtypec #3 \arrowtyped #4
\width #5 \height #6 }
\def\setpos(#1,#2){\xpos=#1 \ypos#2}

\def\settriparms[#1`#2`#3;#4]{\settripairparms[#1`#2`#3`1`1;#4]}%

\def\settripairparms[#1`#2`#3`#4`#5;#6]{%
\arrowtypea #1 \arrowtypeb #2 \arrowtypec #3 \arrowtyped #4
\arrowtypee #5 \width #6 \height #6 }

\def\resetparms{\settripairparms[1`1`1`1`1;500]\width 500}

\resetparms

\def\mvector(#1,#2)#3{
\put(0,0){\vector(#1,#2){#3}}%
\put(0,0){\vector(#1,#2){26}}%
}
\def\evector(#1,#2)#3{{
\arrowlength #3
\put(0,0){\vector(#1,#2){\arrowlength}}%
\advance \arrowlength by-30
\put(0,0){\vector(#1,#2){\arrowlength}}%
}}

\def\horsize#1#2{%
\settowidth{\tempdimen}{$#2$}%
#1=\tempdimen \divide #1 by\unitlength }

\def\vertsize#1#2{%
\settoheight{\tempdimen}{$#2$}%
#1=\tempdimen
\settodepth{\tempdimen}{$#2$}%
\advance #1 by\tempdimen \divide #1 by\unitlength }

\def\putvector(#1,#2)(#3,#4)#5#6{{%
\ifnum3<\arrowtype \putdashvector(#1,#2)(#3,#4)#5\arrowtype \else
\ifnum\arrowtype<-3 \putdashvector(#1,#2)(#3,#4)#5\arrowtype \else
\xpos=#1 \ypos=#2 \run=#3 \rise=#4 \arrowlength=#5 \ifnum
\arrowtype<0
    \ifnum \run=0
        \advance \ypos by-\arrowlength
    \else
        \tempcounta \arrowlength
        \multiply \tempcounta by\rise
        \divide \tempcounta by\run
        \ifnum\run>0
            \advance \xpos by\arrowlength
            \advance \ypos by\tempcounta
        \else
            \advance \xpos by-\arrowlength
            \advance \ypos by-\tempcounta
        \fi
    \fi
    \multiply \arrowtype by-1
    \multiply \rise by-1
    \multiply \run by-1
\fi \ifcase \arrowtype
\or \put(\xpos,\ypos){\vector(\run,\rise){\arrowlength}}%
\or \put(\xpos,\ypos){\mvector(\run,\rise)\arrowlength}%
\or \put(\xpos,\ypos){\evector(\run,\rise){\arrowlength}}%
\fi\fi\fi }}

\def\putsplitvector(#1,#2)#3#4{
\xpos #1 \ypos #2 \arrowtype #4 \halflength #3 \arrowlength #3
\gap 140 \advance \halflength by-\gap \divide \halflength by2
\ifnum\arrowtype>0
   \ifcase \arrowtype
   \or \put(\xpos,\ypos){\line(0,-1){\halflength}}%
       \advance\ypos by-\halflength
       \advance\ypos by-\gap
       \put(\xpos,\ypos){\vector(0,-1){\halflength}}%
   \or \put(\xpos,\ypos){\line(0,-1)\halflength}%
       \put(\xpos,\ypos){\vector(0,-1)3}%
       \advance\ypos by-\halflength
       \advance\ypos by-\gap
       \put(\xpos,\ypos){\vector(0,-1){\halflength}}%
   \or \put(\xpos,\ypos){\line(0,-1)\halflength}%
       \advance\ypos by-\halflength
       \advance\ypos by-\gap
       \put(\xpos,\ypos){\evector(0,-1){\halflength}}%
   \fi
\else \arrowtype=-\arrowtype
   \ifcase\arrowtype
   \or \advance \ypos by-\arrowlength
       \put(\xpos,\ypos){\line(0,1){\halflength}}%
       \advance\ypos by\halflength
       \advance\ypos by\gap
       \put(\xpos,\ypos){\vector(0,1){\halflength}}%
   \or \advance \ypos by-\arrowlength
       \put(\xpos,\ypos){\line(0,1)\halflength}%
       \put(\xpos,\ypos){\vector(0,1)3}%
       \advance\ypos by\halflength
       \advance\ypos by\gap
       \put(\xpos,\ypos){\vector(0,1){\halflength}}%
   \or \advance \ypos by-\arrowlength
       \put(\xpos,\ypos){\line(0,1)\halflength}%
       \advance\ypos by\halflength
       \advance\ypos by\gap
       \put(\xpos,\ypos){\evector(0,1){\halflength}}%
   \fi
\fi }

\def\putmorphism(#1)(#2,#3)[#4`#5`#6]#7#8#9{{%
\run #2 \rise #3 \ifnum\rise=0
  \puthmorphism(#1)[#4`#5`#6]{#7}{#8}#9%
\else\ifnum\run=0
  \putvmorphism(#1)[#4`#5`#6]{#7}{#8}#9%
\else
\setpos(#1)%
\arrowlength #7 \arrowtype #8 \ifnum\run=0 \else\ifnum\rise=0
\else \ifnum\run>0
    \coefa=1
\else
   \coefa=-1
\fi \ifnum\arrowtype>0
   \coefb=0
   \coefc=-1
\else
   \coefb=\coefa
   \coefc=1
   \arrowtype=-\arrowtype
\fi \width=2 \multiply \width by\run \divide \width by\rise
\ifnum \width<0  \width=-\width\fi \advance\width by60 \if l#9
\width=-\width\fi
\putbox(\xpos,\ypos){#4}
{\multiply \coefa by\arrowlength
\advance\xpos by\coefa \multiply \coefa by\rise \divide \coefa
by\run \advance \ypos by\coefa
\putbox(\xpos,\ypos){#5} }%
{\multiply \coefa by\arrowlength
\divide \coefa by2 \advance \xpos by\coefa \advance \xpos by\width
\multiply \coefa by\rise \divide \coefa by\run \advance \ypos
by\coefa
\if l#9%
   \putrbox(\xpos,\ypos){#6}%
\else\if r#9%
   \putlbox(\xpos,\ypos){#6}%
\fi\fi }%
{\multiply \rise by-\coefc
\multiply \run by-\coefc \multiply \coefb by\arrowlength \advance
\xpos by\coefb \multiply \coefb by\rise \divide \coefb by\run
\advance \ypos by\coefb \multiply \coefc by70 \advance \ypos
by\coefc \multiply \coefc by\run \divide \coefc by\rise \advance
\xpos by\coefc \multiply \coefa by140 \multiply \coefa by\run
\divide \coefa by\rise \advance \arrowlength by\coefa
\ifcase\arrowtype
\or \put(\xpos,\ypos){\vector(\run,\rise){\arrowlength}}%
\or \put(\xpos,\ypos){\mvector(\run,\rise){\arrowlength}}%
\or \put(\xpos,\ypos){\evector(\run,\rise){\arrowlength}}%
\fi}\fi\fi\fi\fi}}

\newcount\numbdashes \newcount\lengthdash \newcount\increment

\def\howmanydashes{
\numbdashes=\arrowlength \lengthdash=40 \divide\numbdashes by
\lengthdash \lengthdash=\arrowlength \divide\lengthdash by
\numbdashes
\increment=\lengthdash \multiply\lengthdash by 3
\divide\lengthdash by 5 }

\def\putdashvector(#1)(#2,#3)#4#5{%
\ifnum#3=0 \putdashhvector(#1){#4}#5 \else \ifnum#2=0
\putdashvvector(#1){#4}#5\fi\fi}

\def\putdashhvector(#1,#2)#3#4{{%
\arrowlength=#3 \howmanydashes
\multiput(#1,#2)(\increment,0){\numbdashes}%
{\vrule height .4pt width \lengthdash\unitlength} \arrowtype=#4
\xpos=#1 \ifnum\arrowtype<0 \advance\arrowtype by 7 \fi
\ifcase\arrowtype \or \advance\xpos by 10
    \put(\xpos,#2){\vector(-1,0){\lengthdash}}
    \advance\xpos by 40
    \put(\xpos,#2){\vector(-1,0){\lengthdash}}
\or \advance \xpos by 10
    \put(\xpos,#2){\vector(-1,0){\lengthdash}}
    \advance\xpos by  \arrowlength
    \advance\xpos by  -50
    \put(\xpos,#2){\vector(-1,0){\lengthdash}}
\or \advance\xpos by 10
    \put(\xpos,#2){\vector(-1,0){\lengthdash}}
\or \advance\xpos by \arrowlength
    \advance\xpos by -\lengthdash
    \put(\xpos,#2){\vector(1,0){\lengthdash}}
\or {\advance\xpos by 10
    \put(\xpos,#2){\vector(1,0){\lengthdash}}}
    \advance\xpos by \arrowlength
    \advance\xpos by -\lengthdash
    \put(\xpos,#2){\vector(1,0){\lengthdash}}
\or \advance\xpos by \arrowlength
    \advance\xpos by -\lengthdash
    \put(\xpos,#2){\vector(1,0){\lengthdash}}
    \advance\xpos by -40
    \put(\xpos,#2){\vector(1,0){\lengthdash}}
   \fi
}}

\def\putdashvvector(#1,#2)#3#4{{%
\arrowlength=#3 \howmanydashes \ypos=#2 \advance\ypos by
-\arrowlength
\multiput(#1,#2)(0,\increment){\numbdashes}%
    {\vrule width .4pt height \lengthdash\unitlength}
\arrowtype=#4 \ypos=#2 \ifnum\arrowtype<0 \advance\arrowtype by 7
\fi \ifcase\arrowtype \or \advance\ypos by \arrowlength
\advance\ypos by -40
    \put(#1,\ypos){\vector(0,1){\lengthdash}}
    \advance\ypos by -40
    \put(#1,\ypos){\vector(0,1){\lengthdash}}
\or \advance\ypos by 10
    \put(#1,\ypos){\vector(0,1){\lengthdash}}
    \advance\ypos by \arrowlength \advance\ypos by -40
    \put(#1,\ypos){\vector(0,1){\lengthdash}}
\or \advance\ypos by \arrowlength \advance\ypos by -40
    \put(#1,\ypos){\vector(0,1){\lengthdash}}
\or \advance\ypos by 10
    \put(#1,\ypos){\vector(0,-1){\lengthdash}}
\or \advance\ypos by 10
    \put(#1,\ypos){\vector(0,-1){\lengthdash}}
    \advance\ypos by \arrowlength \advance\ypos by -40
    \put(#1,\ypos){\vector(0,-1){\lengthdash}}
\or \advance\ypos by 10
    \put(#1,\ypos){\vector(0,-1){\lengthdash}}
    \advance\ypos by 40
    \put(#1,\ypos){\vector(0,-1){\lengthdash}}
\fi }}

\def\puthmorphism(#1,#2)[#3`#4`#5]#6#7#8{{%
\xpos #1 \ypos #2 \width #6 \arrowlength #6 \arrowtype=#7
\putbox(\xpos,\ypos){#3\vphantom{#4}}%
{\advance \xpos by\arrowlength
\putbox(\xpos,\ypos){\vphantom{#3}#4}}%
\horsize{\tempcounta}{#3}%
\horsize{\tempcountb}{#4}%
\divide \tempcounta by2 \divide \tempcountb by2 \advance
\tempcounta by30 \advance \tempcountb by30 \advance \xpos
by\tempcounta \advance \arrowlength by-\tempcounta \advance
\arrowlength by-\tempcountb
\putvector(\xpos,\ypos)(1,0)\arrowlength\arrowtype \divide
\arrowlength by2 \advance \xpos by\arrowlength
\vertsize{\tempcounta}{#5}%
\divide\tempcounta by2 \advance \tempcounta by20
\if a#8 %
   \advance \ypos by\tempcounta
   \putbox(\xpos,\ypos){#5}%
\else
   \advance \ypos by-\tempcounta
   \putbox(\xpos,\ypos){#5}%
\fi}}

\def\putvmorphism(#1,#2)[#3`#4`#5]#6#7#8{{%
\xpos #1 \ypos #2 \arrowlength #6 \arrowtype #7
\settowidth{\xlen}{$#5$}%
\putbox(\xpos,\ypos){#3}%
{\advance \ypos by-\arrowlength
\putbox(\xpos,\ypos){#4}}%
{\advance\arrowlength by-140 \advance \ypos by-70 \ifdim\xlen>0pt
   \if m#8%
      \putsplitvector(\xpos,\ypos)\arrowlength\arrowtype
   \else
   \putvector(\xpos,\ypos)(0,-1)\arrowlength\arrowtype
   \fi
\else
   \putvector(\xpos,\ypos)(0,-1)\arrowlength\arrowtype
\fi}%
\ifdim\xlen>0pt
   \divide \arrowlength by2
   \advance\ypos by-\arrowlength
   \if l#8%
      \advance \xpos by-40
      \putrbox(\xpos,\ypos){#5}%
   \else\if r#8%
      \advance \xpos by40
      \putlbox(\xpos,\ypos){#5}%
   \else
      \putbox(\xpos,\ypos){#5}%
   \fi\fi
\fi }}

\def\putsquarep<#1>(#2)[#3;#4`#5`#6`#7]{{%
\setsqparms[#1]%
\setpos(#2)%
\settokens`#3`%
\puthmorphism(\xpos,\ypos)[\tokenc`\tokend`{#7}]{\width}{\arrowtyped}b%
\advance\ypos by \height
\puthmorphism(\xpos,\ypos)[\tokena`\tokenb`{#4}]{\width}{\arrowtypea}a%
\putvmorphism(\xpos,\ypos)[``{#5}]{\height}{\arrowtypeb}l%
\advance\xpos by \width
\putvmorphism(\xpos,\ypos)[``{#6}]{\height}{\arrowtypec}r%
}}

\def\putsquare{\@ifnextchar <{\putsquarep}{\putsquarep%
   <\arrowtypea`\arrowtypeb`\arrowtypec`\arrowtyped;\width`\height>}}
\def\square{\@ifnextchar< {\squarep}{\squarep
   <\arrowtypea`\arrowtypeb`\arrowtypec`\arrowtyped;\width`\height>}}
\def\squarep<#1>[#2`#3`#4`#5;#6`#7`#8`#9]{{
\setsqparms[#1]
\diagram
\putsquarep<\arrowtypea`\arrowtypeb`\arrowtypec`
\arrowtyped;\width`\height>
(0,0)[#2`#3`#4`{#5};#6`#7`#8`{#9}]
\enddiagram
}}                                                 
\def\putptrianglep<#1>(#2,#3)[#4`#5`#6;#7`#8`#9]{{%
\settriparms[#1]%
\xpos=#2 \ypos=#3 \advance\ypos by \height
\puthmorphism(\xpos,\ypos)[#4`#5`{#7}]{\height}{\arrowtypea}a%
\putvmorphism(\xpos,\ypos)[`#6`{#8}]{\height}{\arrowtypeb}l%
\advance\xpos by\height
\putmorphism(\xpos,\ypos)(-1,-1)[``{#9}]{\height}{\arrowtypec}r%
}}

\def\putptriangle{\@ifnextchar <{\putptrianglep}{\putptrianglep
   <\arrowtypea`\arrowtypeb`\arrowtypec;\height>}}
\def\ptriangle{\@ifnextchar <{\ptrianglep}{\ptrianglep
   <\arrowtypea`\arrowtypeb`\arrowtypec;\height>}}
\def\ptrianglep<#1>[#2`#3`#4;#5`#6`#7]{{
\settriparms[#1]
\diagram
\putptrianglep<\arrowtypea`\arrowtypeb`
\arrowtypec;\height>
(0,0)[#2`#3`#4;#5`#6`{#7}]
\enddiagram
}}                                            

\def\putqtrianglep<#1>(#2,#3)[#4`#5`#6;#7`#8`#9]{{%
\settriparms[#1]%
\xpos=#2 \ypos=#3 \advance\ypos by\height
\puthmorphism(\xpos,\ypos)[#4`#5`{#7}]{\height}{\arrowtypea}a%
\putmorphism(\xpos,\ypos)(1,-1)[``{#8}]{\height}{\arrowtypeb}l%
\advance\xpos by\height
\putvmorphism(\xpos,\ypos)[`#6`{#9}]{\height}{\arrowtypec}r%
}}

\def\putqtriangle{\@ifnextchar <{\putqtrianglep}{\putqtrianglep
   <\arrowtypea`\arrowtypeb`\arrowtypec;\height>}}
\def\qtriangle{\@ifnextchar <{\qtrianglep}{\qtrianglep
   <\arrowtypea`\arrowtypeb`\arrowtypec;\height>}}
\def\qtrianglep<#1>[#2`#3`#4;#5`#6`#7]{{
\settriparms[#1]
\width=\height                                
\diagram
\putqtrianglep<\arrowtypea`\arrowtypeb`
\arrowtypec;\height>
(0,0)[#2`#3`#4;#5`#6`{#7}]
\enddiagram
}}

\def\putdtrianglep<#1>(#2,#3)[#4`#5`#6;#7`#8`#9]{{%
\settriparms[#1]%
\xpos=#2 \ypos=#3
\puthmorphism(\xpos,\ypos)[#5`#6`{#9}]{\height}{\arrowtypec}b%
\advance\xpos by \height \advance\ypos by\height
\putmorphism(\xpos,\ypos)(-1,-1)[``{#7}]{\height}{\arrowtypea}l%
\putvmorphism(\xpos,\ypos)[#4``{#8}]{\height}{\arrowtypeb}r%
}}

\def\putdtriangle{\@ifnextchar <{\putdtrianglep}{\putdtrianglep
   <\arrowtypea`\arrowtypeb`\arrowtypec;\height>}}
\def\dtriangle{\@ifnextchar <{\dtrianglep}{\dtrianglep
   <\arrowtypea`\arrowtypeb`\arrowtypec;\height>}}
\def\dtrianglep<#1>[#2`#3`#4;#5`#6`#7]{{
\settriparms[#1]
\width=\height                                
\diagram
\putdtrianglep<\arrowtypea`\arrowtypeb`
\arrowtypec;\height>
(0,0)[#2`#3`#4;#5`#6`{#7}]
\enddiagram
}}

\def\putbtrianglep<#1>(#2,#3)[#4`#5`#6;#7`#8`#9]{{%
\settriparms[#1]%
\xpos=#2 \ypos=#3
\puthmorphism(\xpos,\ypos)[#5`#6`{#9}]{\height}{\arrowtypec}b%
\advance\ypos by\height
\putmorphism(\xpos,\ypos)(1,-1)[``{#8}]{\height}{\arrowtypeb}r%
\putvmorphism(\xpos,\ypos)[#4``{#7}]{\height}{\arrowtypea}l%
}}

\def\putbtriangle{\@ifnextchar <{\putbtrianglep}{\putbtrianglep
   <\arrowtypea`\arrowtypeb`\arrowtypec;\height>}}
\def\btriangle{\@ifnextchar <{\btrianglep}{\btrianglep
   <\arrowtypea`\arrowtypeb`\arrowtypec;\height>}}
\def\btrianglep<#1>[#2`#3`#4;#5`#6`#7]{{
\settriparms[#1]
\width=\height                               
\diagram
\putbtrianglep<\arrowtypea`\arrowtypeb`
\arrowtypec;\height>
(0,0)[#2`#3`#4;#5`#6`{#7}]
\enddiagram
}}

\def\putAtrianglep<#1>(#2,#3)[#4`#5`#6;#7`#8`#9]{{%
\settriparms[#1]%
\xpos=#2 \ypos=#3 {\multiply \height by2
\puthmorphism(\xpos,\ypos)[#5`#6`{#9}]{\height}{\arrowtypec}b}%
\advance\xpos by\height \advance\ypos by\height
\putmorphism(\xpos,\ypos)(-1,-1)[#4``{#7}]{\height}{\arrowtypea}l%
\putmorphism(\xpos,\ypos)(1,-1)[``{#8}]{\height}{\arrowtypeb}r%
}}

\def\putAtriangle{\@ifnextchar <{\putAtrianglep}{\putAtrianglep
   <\arrowtypea`\arrowtypeb`\arrowtypec;\height>}}
\def\Atriangle{\@ifnextchar <{\Atrianglep}{\Atrianglep
   <\arrowtypea`\arrowtypeb`\arrowtypec;\height>}}
\def\Atrianglep<#1>[#2`#3`#4;#5`#6`#7]{{
\settriparms[#1]
\width=\height                                     
\diagram
\putAtrianglep<\arrowtypea`\arrowtypeb`
\arrowtypec;\height>
(0,0)[#2`#3`#4;#5`#6`{#7}]
\enddiagram
}}

\def\putAtrianglepairp<#1>(#2)[#3;#4`#5`#6`#7`#8]{{%
\settripairparms[#1]%
\setpos(#2)%
\settokens`#3`%
\puthmorphism(\xpos,\ypos)[\tokenb`\tokenc`{#7}]{\height}{\arrowtyped}b%
\advance\xpos by\height
\puthmorphism(\xpos,\ypos)[\phantom{\tokenc}`\tokend`{#8}]%
{\height}{\arrowtypee}b%
\advance\ypos by\height
\putmorphism(\xpos,\ypos)(-1,-1)[\tokena``{#4}]{\height}{\arrowtypea}l%
\putvmorphism(\xpos,\ypos)[``{#5}]{\height}{\arrowtypeb}m%
\putmorphism(\xpos,\ypos)(1,-1)[``{#6}]{\height}{\arrowtypec}r%
}}

\def\putAtrianglepair{\@ifnextchar <{\putAtrianglepairp}{\putAtrianglepairp%
   <\arrowtypea`\arrowtypeb`\arrowtypec`\arrowtyped`\arrowtypee;\height>}}
\def\Atrianglepair{\@ifnextchar <{\Atrianglepairp}{\Atrianglepairp%
   <\arrowtypea`\arrowtypeb`\arrowtypec`\arrowtyped`\arrowtypee;\height>}}

\def\Atrianglepairp<#1>[#2;#3`#4`#5`#6`#7]{{
\settripairparms[#1]
\settokens`#2`
\width=\height                                
\diagram
\putAtrianglepairp                            
<\arrowtypea`\arrowtypeb`\arrowtypec`
\arrowtyped`\arrowtypee;\height>
(0,0)[{#2};#3`#4`#5`#6`{#7}]
\enddiagram
}}

\def\putVtrianglep<#1>(#2,#3)[#4`#5`#6;#7`#8`#9]{{%
\settriparms[#1]%
\xpos=#2 \ypos=#3 \advance\ypos by\height {\multiply\height by2
\puthmorphism(\xpos,\ypos)[#4`#5`{#7}]{\height}{\arrowtypea}a}%
\putmorphism(\xpos,\ypos)(1,-1)[`#6`{#8}]{\height}{\arrowtypeb}l%
\advance\xpos by\height \advance\xpos by\height
\putmorphism(\xpos,\ypos)(-1,-1)[``{#9}]{\height}{\arrowtypec}r%
}}

\def\putVtriangle{\@ifnextchar <{\putVtrianglep}{\putVtrianglep
   <\arrowtypea`\arrowtypeb`\arrowtypec;\height>}}
\def\Vtriangle{\@ifnextchar <{\Vtrianglep}{\Vtrianglep
   <\arrowtypea`\arrowtypeb`\arrowtypec;\height>}}
\def\Vtrianglep<#1>[#2`#3`#4;#5`#6`#7]{{
\settriparms[#1]
\width=\height                                 
\diagram
\putVtrianglep<\arrowtypea`\arrowtypeb`
\arrowtypec;\height>
(0,0)[#2`#3`#4;#5`#6`{#7}]
\enddiagram
}}

\def\putVtrianglepairp<#1>(#2)[#3;#4`#5`#6`#7`#8]{{
\settripairparms[#1]%
\setpos(#2)%
\settokens`#3`%
\advance\ypos by\height
\putmorphism(\xpos,\ypos)(1,-1)[`\tokend`{#6}]{\height}{\arrowtypec}l%
\puthmorphism(\xpos,\ypos)[\tokena`\tokenb`{#4}]{\height}{\arrowtypea}a%
\advance\xpos by\height
\puthmorphism(\xpos,\ypos)[\phantom{\tokenb}`\tokenc`{#5}]%
{\height}{\arrowtypeb}a%
\putvmorphism(\xpos,\ypos)[``{#7}]{\height}{\arrowtyped}m%
\advance\xpos by\height
\putmorphism(\xpos,\ypos)(-1,-1)[``{#8}]{\height}{\arrowtypee}r%
}}

\def\putVtrianglepair{\@ifnextchar <{\putVtrianglepairp}{\putVtrianglepairp%
    <\arrowtypea`\arrowtypeb`\arrowtypec`\arrowtyped`\arrowtypee;\height>}}
\def\Vtrianglepair{\@ifnextchar <{\Vtrianglepairp}{\Vtrianglepairp%
    <\arrowtypea`\arrowtypeb`\arrowtypec`\arrowtyped`\arrowtypee;\height>}}
\def\Vtrianglepairp<#1>[#2;#3`#4`#5`#6`#7]{{
\settripairparms[#1]
\settokens`#2`
\diagram
\putVtrianglepairp                             
<\arrowtypea`\arrowtypeb`\arrowtypec`
\arrowtyped`\arrowtypee;\height>
(0,0)[{#2};#3`#4`#5`#6`{#7}]
\enddiagram
}}

\def\putCtrianglep<#1>(#2,#3)[#4`#5`#6;#7`#8`#9]{{%
\settriparms[#1]%
\xpos=#2 \ypos=#3 \advance\ypos by\height
\putmorphism(\xpos,\ypos)(1,-1)[``{#9}]{\height}{\arrowtypec}l%
\advance\xpos by\height \advance\ypos by\height
\putmorphism(\xpos,\ypos)(-1,-1)[#4`#5`{#7}]{\height}{\arrowtypea}l%
{\multiply\height by 2
\putvmorphism(\xpos,\ypos)[`#6`{#8}]{\height}{\arrowtypeb}r}%
}}

\def\putCtriangle{\@ifnextchar <{\putCtrianglep}{\putCtrianglep
    <\arrowtypea`\arrowtypeb`\arrowtypec;\height>}}
\def\Ctriangle{\@ifnextchar <{\Ctrianglep}{\Ctrianglep
    <\arrowtypea`\arrowtypeb`\arrowtypec;\height>}}
\def\Ctrianglep<#1>[#2`#3`#4;#5`#6`#7]{{
\settriparms[#1]
\width=\height                               
\diagram
\putCtrianglep<\arrowtypea`\arrowtypeb`
\arrowtypec;\height>
(0,0)[#2`#3`#4;#5`#6`{#7}]
\enddiagram
}}                                           
\def\putDtrianglep<#1>(#2,#3)[#4`#5`#6;#7`#8`#9]{{%
\settriparms[#1]%
\xpos=#2 \ypos=#3 \advance\xpos by\height \advance\ypos by\height
\putmorphism(\xpos,\ypos)(-1,-1)[``{#9}]{\height}{\arrowtypec}r%
\advance\xpos by-\height \advance\ypos by\height
\putmorphism(\xpos,\ypos)(1,-1)[`#5`{#8}]{\height}{\arrowtypeb}r%
{\multiply\height by 2
\putvmorphism(\xpos,\ypos)[#4`#6`{#7}]{\height}{\arrowtypea}l}%
}}

\def\putDtriangle{\@ifnextchar <{\putDtrianglep}{\putDtrianglep
    <\arrowtypea`\arrowtypeb`\arrowtypec;\height>}}
\def\Dtriangle{\@ifnextchar <{\Dtrianglep}{\Dtrianglep
   <\arrowtypea`\arrowtypeb`\arrowtypec;\height>}}
\def\Dtrianglep<#1>[#2`#3`#4;#5`#6`#7]{{
\settriparms[#1]
\width=\height                              
\diagram
\putDtrianglep<\arrowtypea`\arrowtypeb`
\arrowtypec;\height>
(0,0)[#2`#3`#4;#5`#6`{#7}]
\enddiagram
}}                                          
\def\setrecparms[#1`#2]{\width=#1 \height=#2}%

\def\recursep<#1`#2>[#3;#4`#5`#6`#7`#8]{{\m@th
\width=#1 \height=#2 \settokens`#3`
\settowidth{\tempdimen}{$\tokena$} \ifdim\tempdimen=0pt
  \savebox{\tempboxa}{\hbox{$\tokenb$}}%
  \savebox{\tempboxb}{\hbox{$\tokend$}}%
  \savebox{\tempboxc}{\hbox{$#6$}}%
\else
  \savebox{\tempboxa}{\hbox{$\hbox{$\tokena$}\times\hbox{$\tokenb$}$}}%
  \savebox{\tempboxb}{\hbox{$\hbox{$\tokena$}\times\hbox{$\tokend$}$}}%
  \savebox{\tempboxc}{\hbox{$\hbox{$\tokena$}\times\hbox{$#6$}$}}%
\fi \ypos=\height \divide\ypos by 2 \xpos=\ypos \advance\xpos by
\width \bfig
\putCtrianglep<-1`1`1;\ypos>(0,0)[`\tokenc`;#5`#6`{#7}]%
\puthmorphism(\ypos,0)[\tokend`\usebox{\tempboxb}`{#8}]{\width}{-1}b%
\puthmorphism(\ypos,\height)[\tokenb`\usebox{\tempboxa}`{#4}]{\width}{-1}a%
\advance\ypos by \width
\putvmorphism(\ypos,\height)[``\usebox{\tempboxc}]{\height}1r%
\efig }}

\def\recurse{\@ifnextchar <{\recursep}{\recursep<\width`\height>}}

\def\puttwohmorphisms(#1,#2)[#3`#4;#5`#6]#7#8#9{{%
\puthmorphism(#1,#2)[#3`#4`]{#7}0a \ypos=#2 \advance\ypos by 20
\puthmorphism(#1,\ypos)[\phantom{#3}`\phantom{#4}`#5]{#7}{#8}a
\advance\ypos by -40
\puthmorphism(#1,\ypos)[\phantom{#3}`\phantom{#4}`#6]{#7}{#9}b }}

\def\puttwovmorphisms(#1,#2)[#3`#4;#5`#6]#7#8#9{{%
\putvmorphism(#1,#2)[#3`#4`]{#7}0a \xpos=#1 \advance\xpos by -20
\putvmorphism(\xpos,#2)[\phantom{#3}`\phantom{#4}`#5]{#7}{#8}l
\advance\xpos by 40
\putvmorphism(\xpos,#2)[\phantom{#3}`\phantom{#4}`#6]{#7}{#9}r }}

\def\puthcoequalizer(#1)[#2`#3`#4;#5`#6`#7]#8#9{{%
\setpos(#1)%
\puttwohmorphisms(\xpos,\ypos)[#2`#3;#5`#6]{#8}11%
\advance\xpos by #8
\puthmorphism(\xpos,\ypos)[\phantom{#3}`#4`#7]{#8}1{#9} }}

\def\putvcoequalizer(#1)[#2`#3`#4;#5`#6`#7]#8#9{{%
\setpos(#1)%
\puttwovmorphisms(\xpos,\ypos)[#2`#3;#5`#6]{#8}11%
\advance\ypos by -#8
\putvmorphism(\xpos,\ypos)[\phantom{#3}`#4`#7]{#8}1{#9} }}

\def\putthreehmorphisms(#1)[#2`#3;#4`#5`#6]#7(#8)#9{{%
\setpos(#1) \settypes(#8)
\if a#9 %
     \vertsize{\tempcounta}{#5}%
     \vertsize{\tempcountb}{#6}%
     \ifnum \tempcounta<\tempcountb \tempcounta=\tempcountb \fi
\else
     \vertsize{\tempcounta}{#4}%
     \vertsize{\tempcountb}{#5}%
     \ifnum \tempcounta<\tempcountb \tempcounta=\tempcountb \fi
\fi \advance \tempcounta by 60
\puthmorphism(\xpos,\ypos)[#2`#3`#5]{#7}{\arrowtypeb}{#9}
\advance\ypos by \tempcounta
\puthmorphism(\xpos,\ypos)[\phantom{#2}`\phantom{#3}`#4]{#7}{\arrowtypea}{#9}
\advance\ypos by -\tempcounta \advance\ypos by -\tempcounta
\puthmorphism(\xpos,\ypos)[\phantom{#2}`\phantom{#3}`#6]{#7}{\arrowtypec}{#9}
}}

\def\setarrowtoks[#1`#2`#3`#4`#5`#6]{%
\def\toka{#1}
\def\tokb{#2}
\def\tokc{#3}
\def\tokd{#4}
\def\toke{#5}
\def\tokf{#6}
}
\def\hex{\@ifnextchar <{\hexp}{\hexp<1000`400>}}
\def\hexp<#1`#2>[#3`#4`#5`#6`#7`#8;#9]{%
\setarrowtoks[#9] \yext=#2 \advance \yext by #2 \xext=#1
\advance\xext by \yext \bfig
\putCtriangle<-1`0`1;#2>(0,0)[`#5`;\tokb``\tokd] \xext=#1
\yext=#2 \advance \yext by #2
\putsquare<1`0`0`1;\xext`\yext>(#2,0)[#3`#4`#7`#8;\toka```\tokf]
\advance \xext by #2
\putDtriangle<0`1`-1;#2>(\xext,0)[`#6`;`\tokc`\toke] \efig }

\makeatother

\begin{document}
\begin{frontmatter}

\title{Life--Space Foam: a Medium for\\ Motivational and Cognitive Dynamics}
\author{Vladimir Ivancevic and Eugene Aidman}
\address{Human Systems Integration, Land Operations Division\\
Defence Science \& Technology Organisation, AUSTRALIA}

\begin{abstract}
General stochastic dynamics, developed in a framework of Feynman
path integrals, have been applied to Lewinian field--theoretic
psychodynamics \cite{Lewin51,Lewin97,Gold}, resulting in the
development of a new concept of \emph{life--space foam} (LSF) as a
natural medium for motivational and cognitive psychodynamics.
According to LSF formalisms, the classic Lewinian life space can
be macroscopically represented as a smooth manifold with steady
force--fields and behavioral paths, while at the microscopic level
it is more realistically represented as a collection of wildly
fluctuating force--fields, (loco)motion paths and local geometries
(and topologies with holes). A set of least--action principles is
used to model the smoothness of global, macro--level LSF paths,
fields and geometry. To model the corresponding local,
micro--level LSF structures, an adaptive path integral is used,
defining a multi--phase and multi--path (multi--field and
multi--geometry) transition process from intention to goal--driven
action. Application examples of this new approach include (but are
not limited to) information processing, motivational fatigue,
learning, memory and decision--making.\bigbreak


\end{abstract}

\end{frontmatter}

\section{Introduction}

One of the key challenges in modelling complex human behavior is
combining enough detail with sufficient scope in a given
representation. A trade--off between detail and scope seems
inevitable: individual models are either broad or detailed -- but
almost never both. However, most recent advances in understanding
complex biopsychosocial phenomena have been associated with
multilevel approaches, such as ascertaining the genetic
contribution to developmental psychopathology through interactive
analysis of ``precisely
measured microenvironments and well--specified macroenvironments" \cite%
{Hinshaw}. The present article develops a formalism intended to
capture the complexity of motivated human behavior without
sacrificing either detail or scope. We will discuss nonlinear
stochastic methods (as a generalization of conventional
statistics) leading to a more sophisticated conception of quantum
probability in a framework of Feynmanian path integral
applications. Lewin's force--field theory
\cite{Lewin51,Lewin97,Gold} will be used to illustrate how the new
approach can be applied.

Applications of Nonlinear Dynamic Systems (NDS) theory in
psychology have been encouraging, if not universally effective
\cite{Metzger}. Its historical antecedents can be traced back to
Piaget's \cite{Piaget} and Vygotsky's \cite{Vygotsky}
interpretations of the dynamic relations between action and
thought, Lewin's theory of social dynamics and
cognitive--affective development \cite{Lewin97}, and
\cite{Bernstein} theory of self--adjusting, goal--driven motor
action.

Characteristically, one of the most productive applications of NDS
to date
is in the area of motor development (see \cite{Schoener,Smith,Thelen,Turvey}%
), with its subject matter lending itself to NDS treatment by
possessing the required properties of an NDS entity: even a
relatively simple skill, such as walking, is characterized by a
stable, adaptive and self--organizing pattern, different from its
component skills and capable of developing, from a wide range of
starting points, to a well identified, stable attractor (e.g.,
most children, including those with impediments, eventually master
a skill recognizable as \textquotedblleft walking") \cite{Eimas}.
Cognitive
development seems to be characterized by similar dynamical patterns \cite%
{Bogartz,Thelen}. For example, speaking in sentences is
qualitatively different from its component skills of remembering
words and voice production; and most children are eventually
capable of creating sentences \textquotedblleft on the fly",
fluently adapting to the constraints of their native language.

One important conclusion: in order for the powerful formalisms of
NDS to be effective in an application, its subject matter should
possess NDS--relevant properties such as stable, adaptive and
self--organizing patterns of sufficient complexity, non--reducible
to their components and capable of developing to stable
attractors.

In the light of the above requirement, current conceptualizations
of decision making (see \cite{Zsambok,Busemeyer2}) don't seem to
be appropriate for NDS treatment. In particular, in the context of
the two levels of analysis proposed below, most phenomena captured
by the Decision Field Theory (DFT) \cite{Busemeyer}, including its
multi--alternative version (MDFT) \cite{Roe}, occur at a
macroscopic level, while their microscopic underpinnings are
either not known or of little consequence. Decision making, in
essence, is about choice from finite alternatives -- e.g., among
competing courses of action (COA). However, once a certain COA is
chosen (i.e., decided upon), implementing that decision (i.e.,
conducting the chosen action) introduces such levels of
complexity, even in relatively simple environments, that
microscopic level of analysis becomes critical. And that's where
both the classical NDS and its generalization in the
path--integral form, show their advantage.

As a simulation example of this new approach, we propose a
path--integral multialternative decision--field model, as a
generalization to the abstract linear discrete--time stochastic
model of \cite{Roe}, operating on a single level of a linear
multiatribute space. Our path--integral approach proposes a
nonlinear, hybrid (i.e., both continuous and discrete--time),
stochastic sum--over--histories mechanism, based on the
\textit{quantum probability concept}, operating at two distinct
levels of the Lewinian `life space'.

\subsection{Lewinian Life Space}

Both the original \textit{Lewinian force--field theory} in
psychology (see \cite{Lewin51,Lewin97,Gold}) and modern
decision--field dynamics (see \cite{Busemeyer,Roe,Busemeyer2}) are
based on the classical Lewinian concept of an individual's
\textit{life space}. As a topological construct, Lewinian life
space represents a person's psychological environment that
contains \textit{regions} separated by dynamic permeable
\textit{boundaries}. As a field construct, on the other hand, the
life space is not empty: each of its regions is characterized by
\textit{valence} (ranging from positive or negative and resulting
from an interaction between the person's \textit{needs} and the
dynamics of their \textit{environment}). Need is an energy
construct, according to Lewin. It creates \textit{tension} in the
person, which, in combination with other tensions, initiates and
sustains behavior. Needs vary from the most primitive urges to the
most idiosyncratic intentions and can be both internally generated
(e.g., thirst, hunger or sex) and stimulus--induced (e.g., an urge
to buy something in response to a TV advertisement). Valences are,
in essence, personal values dynamically derived from the person's
needs and attached to various regions in their life space. As a
field, the life space generates forces pulling the person towards
positively--valenced regions and pushing them away from regions
with negative valence. Lewin's term for these forces is
\textit{vectors}. Combinations of multiple vectors in the life
space cause the person to move from one region towards another.
This movement is termed \textit{locomotion} and it may range from
overt behavior to cognitive shifts (e.g., between alternatives in
a decision--making process). Locomotion normally results in
crossing the boundaries between regions. When their permeability
is degraded, these boundaries become \textit{barriers} that
restrain locomotion. Life space model, thus, offers a
meta--theoretical language to describe a wide range of behaviors,
from goal--directed action to intrapersonal conflicts and
multi--alternative decision--making.

\begin{figure}[tbh]
\centerline{\includegraphics[width=13cm]{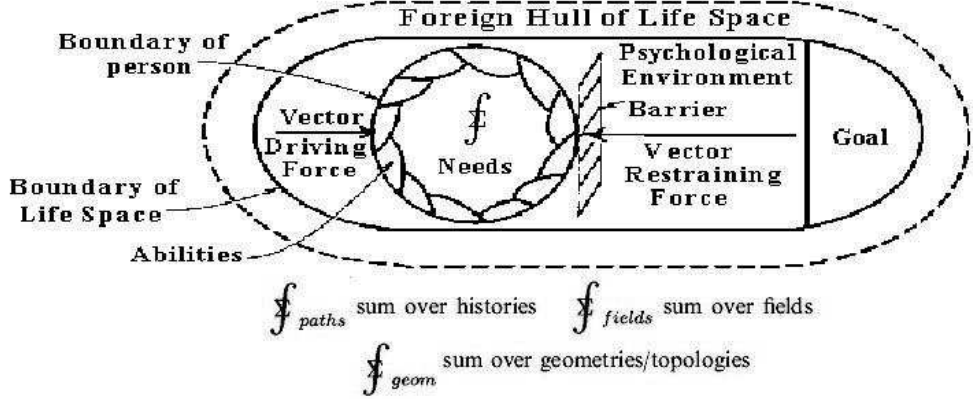}}
\caption{Diagram of the \textit{life space foam}: classical
representation of Lewinian life space, with an adaptive path
integral $\put(0,0){\LARGE $\int$}\put(20,15){\small
$\Sigma$}\quad$ (see footnote 1) acting inside it and generating
microscopic fluctuation dynamics.} \label{LifeSpace}
\end{figure}

In order to formalize the Lewinian life--space concept, a set of
\emph{action principles} need to be associated to Lewinian
force--fields, (loco)motion paths (representing mental
abstractions of biomechanical paths \cite{VladSIAM}) and life
space geometry. As an extension of the Lewinian concept, in this
paper we introduce a new concept of \textit{life--space foam}
(LSF, see Figure \ref{LifeSpace}). According to this new concept,
Lewin's life space can be represented as a geometrical object with
globally smooth macro--dynamics, which is at the same time
underpinned by wildly fluctuating, non--smooth, local
micro--dynamics, describable by sum--over--histories
$\put(0,0){\LARGE $\int$}\put(20,15){\small
$\Sigma$}\quad\,_{paths}\,$, sum--over--fields $\put(0,0){\LARGE
$\int$}\put(20,15){\small $\Sigma$}\quad\,_{fields}$ and
sum--over--geometries/topologies $\put(0,0){\LARGE
$\int$}\put(20,15){\small $\Sigma$}\quad\,_{geom}$ \footnote{We
use the peculiar Dirac's quantum symbol $\put(0,0){\LARGE
$\int$}\put(20,15){\small $\Sigma$}\quad\,$ to denote summation
over `discrete spectrum' and integration over `continuous
spectrum' of paths, fields and geometries in the microscopic level
of the \emph{Lewinian life space}.}.

LSF is thus a two--level geometrodynamical object, representing
these two distinct types of dynamics in the life space. At its
\textit{macroscopic spatio--temporal level}, LSF appears as a
`nice \& smooth' geometrical object with globally predictable
dynamics -- formally, a smooth $n-$dimensional manifold $M$ with
Riemannian metric $g_{ij}$ (compare with \cite{Amari,PMZ}), smooth
force--fields and smooth (loco)motion paths, as conceptualized in
the Lewinian theory. To model the global and smooth macro--level
LSF--paths, fields and geometry, we use the general physics--like
\emph{principle of the least action}.

Now, the apparent smoothness of the macro--level LSF is achieved
by the existence of another level underneath it. This
\textit{micro--level} LSF is actually a collection of wildly
fluctuating force--fields, (loco)motion paths, curved regional
geometries and topologies with holes. The micro--level LSF is
proposed as an extension of the Lewinian concept: it is
characterized by uncertainties and fluctuations, enabled by
microscopic time--level, microscopic transition paths, microscopic
force--fields, local geometries and varying topologies with holes.
To model these fluctuating microscopic LSF--structures, we use
three instances of \textit{adaptive path integral}, defining a
multi--phase and multi--path (also multi--field and
multi--geometry) \textit{transition} process from
\textit{intention} to the goal--driven \textit{action}.

We use the new LSF concept to develop modelling framework for
motivational dynamics (MD) and induced cognitive dynamics (CD).

According to Heckhausen (see \cite{Heckhausen}),
\textit{motivation} can be thought of as a process of
\textit{energizing} and \textit{directing the action}. The process
of energizing can be represented by Lewin's \textit{force--field
analysis} and Vygotsky's \textit{motive formation} (see
\cite{Vygotsky,Aidman}), while the process of directing can be
represented by \textit{hierarchical action control} (see
\cite{Bernstein,Kuhl}).

\begin{figure}[tbh]
\centerline{\includegraphics[width=10cm]{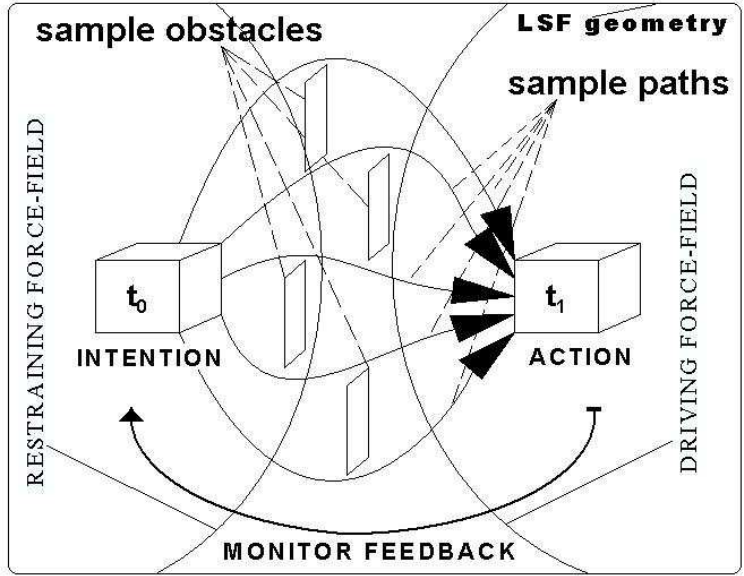}}
\caption{\textit{Transition--propagator} corresponding to each of
the motivational phases
$\{\mathcal{F},\mathcal{I},\mathcal{M},\mathcal{T}\}$, consisting
of an ensemble of feedforward paths propagating through the `wood
of obstacles'. The paths affected by driving and restraining
force--fields, as well as by the local LSF--geometry. Transition
goes from $Intention$, occurring at a sample time instant $t_{0}$,
to $Action$, occurring at some later time $t_{1}$. Each propagator
is controlled by its own $Monitor$ feedback.} \label{IvPaths}
\end{figure}

Motivation processes both precede and coincide with every
goal--directed action. Usually these motivation processes include
the sequence of the following four feedforward \textit{phases}
\cite{Vygotsky,Aidman}: (*)
\begin{enumerate}
    \item \textit{Intention Formation} $\mathcal{F}$, including: decision making, commitment building, etc.
    \item \textit{Action Initiation} $\mathcal{I}$, including: handling conflict of motives, resistance to
    alternatives, etc.
    \item \textit{Maintaining the Action} $\mathcal{M}$, including: resistance to fatigue, distractions, etc.
    \item \textit{Termination} $\mathcal{T}$, including parking and avoiding addiction, i.e., staying in
    control.
\end{enumerate}
With each of the phases
$\{\mathcal{F},\mathcal{I},\mathcal{M},\mathcal{T}\}$ in (*), we
can associate a \textit{transition propagator} -- an ensemble of
(possibly crossing) feedforward paths propagating through the
`wood of obstacles' (including topological holes in the LSF, see
Figure \ref{IvPaths}), so that the complete transition is a
product of propagators (as well as sum over paths). All the
phases--propagators are controlled by a unique $Monitor$ feedback
process.

In this paper we propose an \textit{adaptive path integral}
formulation for these motivational--transitions. In essence, we
sum/integrate over different paths and make a product
(composition) of different phases--propagators. Also, recall that
modern stochastic calculus permits development of three
alternative descriptions of \textit{general Markov stochastic
processes}:\footnote{Recall that \textit{Markov stochastic
process} is a stochastic (random) process characterized by a
\emph{lack of memory}, i.e., the statistical properties of the
immediate future are uniquely determined by the present,
regardless of the past \cite{Gardiner}. This Markov assumption can
be formulated in terms of the conditional probabilities
$P(x^{i},t_{i})$: if the times $t_{i}$ increase from right to
left, the conditional probability is determined entirely by the
knowledge of the most recent condition. The general, continuous +
discrete Markov process is generated by a set of conditional
probabilities whose probability--density evolution,
$P=P(x',t'|x'',t'')$, obeys the general
\textit{Chapman--Kolmogorov integro--differential equation}
\begin{eqnarray*}
\partial _{t}P &=&-\sum_{i}\frac{\partial }{\partial x^{i}}\left\{
A_{i}[x(t),t]\,P\right\} \qquad \\
&+&  \frac{1}{2}\sum_{ij}\frac{\partial ^{2}}{\partial x^{i}}%
\partial {x^{j}}\left\{ B_{ij}[x(t),t]\,P\right\} \qquad\\
&+&  \int dx\left\{ W(x'|x'',t)\,P-W(x''|%
x',t)\,P\right\},
\end{eqnarray*}%
including: \textit{deterministic drift} (the first term on the
right, called the \textit{Liouville equation}), \textit{diffusion
fluctuations} (the second term on the right, called the
\textit{Fokker--Planck equation}) and \textit{discontinuous jumps}
(the third term on the right, called the \textit{Master
equation}).}
\begin{enumerate}
\item Langevin rate equations \cite{Gardiner}, \item
Fokker--Planck equations \cite{Gardiner}, and \item Path integrals
\cite{Graham,Langouche,W-W83a,W-W83b}.
\end{enumerate}
Here we follow the most general, path integral approach, namely it
is the general Chapman--Kolmogorov integro--differential equation,
with its conditional probability density evolution,
$P=P(x',t'|x'',t'')$, that we are going to model by various forms
of the Feynman path integral $\put(0,0){\LARGE
$\int$}\put(20,15){\small $\Sigma$}\quad\,$, providing us with the
physical insight behind the abstract probability densities.

We will also attempt to demonstrate the utility of the same
LSF--formalisms in representing cognitive functions, such as
memory, learning and decision making. For example, in the
classical $Stimulus\,\,encoding\to Search\to Decision\to Response$
sequence \cite{Sternberg,Ashcraft}, the environmental
input--triggered \textit{sensory memory} and \textit{working
memory} (WM) can be interpreted as operating at the micro--level
force--field under the executive control of the $Monitor$
feedback, whereas \textit{search} can be formalized as a
\textit{control} mechanism guiding retrieval from the long--term
memory (LTM, itself shaped by learning) and filtering material
relevant to decision making into the WM. The essential measure of
these mental processes, the \textit{processing speed} (essentially
determined by Sternberg's reaction--time) can be represented by
our (loco)motion speed $\dot{x}$.

\section{Six Facets of the Life Space Foam}

The LSF has three forms of appearance: $paths+fields+geometries$,
acting on both macro--level and micro--level, which is six modes
in total. In this section, we develop three least action
principles for the macro--LSF--level and three adaptive path
integrals for the micro--LSF--level. While developing our
psycho--physical formalism, we will address the behavioral issues
of motivational fatigue, learning, memory and decision making.

\subsection{General Formalism}

At both macro-- and micro--levels, the total LSF represents a
union of transition paths, force--fields and geometries, formally
written as
\begin{equation}
LSF_{total}=LSF_{paths}\bigcup LSF_{fields}\bigcup LSF_{geom}.
\label{LSF}
\end{equation}
Corresponding to each of the three LSF--subspaces in (\ref{LSF})
we formulate:
\begin{enumerate}
    \item The \textit{least action principle}, to model deterministic and
    predictive, macro--level MD and CD, giving a unique, global, causal and smooth path--field--geometry
    on the macroscopic spatio--temporal level; and
    \item Associated \textit{adaptive path integral} to model uncertain, fluctuating and
    probabilistic, micro--level MD and CD, as an ensemble of local paths--fields--geometri-es on
    the microscopic spatio--temporal level, to which the global macro--level MD and CD represents both time and ensemble
    \textit{average} (which are equal according to the \textit{ergodic
    hypothesis}).
\end{enumerate}
In the proposed formalism, transition paths $x^i(t)$ are affected
by the force--fields $\varphi^k(t)$, which are themselves affected
by geometry with metric $g_{ij}$.

\subsubsection{Global Macro--Level of $LSF_{total}$}

In general, at the \textit{macroscopic} LSF--level we first
formulate the \textit{total action} $S[\Phi]$, the central
quantity in our formalism that has psycho--physical dimensions of
$Energy\times Time=Effort$, with immediate cognitive and
motivational applications: \textit{the greater the action -- the
higher the speed of cognitive processes and the lower the
macroscopic fatigue} (which includes all sources of physical,
cognitive and emotional fatigue that influence motivational
dynamics). The action $S[\Phi]$ depends on macroscopic paths,
fields and geometries, commonly denoted by an abstract field
symbol $\Phi^i$. The action $S[\Phi]$ is formally defined as a
temporal integral from the \textit{initial} time instant $t_{ini}$
to the \textit{final} time instant $t_{fin}$,
\begin{equation}
S[\Phi]=\int_{t_{ini}}^{t_{fin}}\mathfrak{L}[\Phi]\,dt,
\label{act}
\end{equation}%
with \textit{Lagrangian density} given by
\begin{equation*}
\mathfrak{L}[\Phi]=\int
d^{n}x\,\mathcal{L}(\Phi_i,\partial_{x^j}\Phi^i),
\end{equation*}%
where the integral is taken over all $n$ coordinates $x^j=x^j(t)$
of the LSF, and $\partial_{x^j}\Phi^i$ are time and space partial
derivatives of the $\Phi^i-$variables over coordinates.

Second, we formulate the \textit{least action principle} as a
minimal variation $\delta$ of the action $S[\Phi]$,
\begin{equation}
\delta S[\Phi]=0, \label{actPr}
\end{equation}
which, using techniques from the calculus of variations gives, in
the form of the so--called Euler--Lagrangian equations, a shortest
(loco)motion path, an extreme force--field, and a life--space
geometry of minimal curvature (and without holes). In this way, we
effectively derive a \textit{unique globally smooth transition
map}
\begin{equation}
F\;:\;INTENTION_{t_{ini}}\cone{} ACTION_{t_{fin}}, \label{unique}
\end{equation}
performed at a macroscopic (global) time--level from some initial
time $t_{ini}$ to the final time $t_{fin}$.

In this way, we get macro--objects in the global LSF: a single
path described by Newtonian--like equation of motion, a single
force--field described by Maxwellian--like field equations, and a
single obstacle--free Riemannian geometry (with global topology
without holes).

For example, in 1945--1949 Wheeler and Feynman developed their
\textit{action--at--a--distance electrodynamics} \cite{FW}, in
complete experimental agreement with the classical Maxwell's
electromagnetic theory, but at the same time avoiding the
complications of divergent self--interaction of the Maxwell's
theory as well as eliminating its infinite number of field degrees
of freedom. In Wheeler--Feynman view, ``Matter consists of
electrically charged particles,'' so they found a form for the
action directly involving the motions of the charges only, which
upon variation would give the Newtonian--like equations of motion
of these charges. Here is the expression for this action in the
flat space--time, which is in the core of quantum electrodynamics:
\begin{eqnarray}
S[x;t_{i},t_{j}] &=&\frac{1}{2}\sum_{i}m_{i}\int (\dot{x}_{\mu
}^{i})^{2}\,dt_{i}+\frac{1}{2}\sum_{i,j}^{i\neq j}e_{i}e_{j}\int
\int \delta (I_{ij}^{2})\,\dot{x}_{\mu }^{i}(t_{i})\dot{x}_{\mu
}^{j}(t_{j})\,dt_{i}dt_{j}  \nonumber \\
&&\text{with}  \label{Fey1} \\
I_{ij}^{2} &=&\left[ x_{\mu }^{i}(t_{i})-x_{\mu
}^{j}(t_{j})\right] \left[ x_{\mu }^{i}(t_{i})-x_{\mu
}^{j}(t_{j})\right] ,  \nonumber
\end{eqnarray}
where $x_{\mu }^{i}=x_{\mu }^{i}(t_{i})$ is the four--vector
position of the $i$th particle as a function of the proper time $t_{i}$, while $%
\dot{x}_{\mu }^{i}(t_{i})=dx_{\mu }^{i}/dt_{i}$ is the velocity
four--vector. The first term in the action (\ref{Fey1}) is the
ordinary mechanical action in Euclidean space, while the second
term defines the electrical interaction of the charges,
representing the Maxwell--like field (it is summed over each pair
of charges; the factor $\frac{1}{2}$ is to count each pair once,
while the term $i=j$ is omitted to avoid self--action; the
interaction is a double integral over a delta function of the
square of space--time interval $I^{2}$ between two points on the
paths; thus, interaction occurs only when this interval vanishes,
that is, along light cones \cite{FW}).

Now, from the point of view of Lewinian geometrical force--fields
and (loco)mo-tion paths, we can give the following life--space
interpretation to the Wheeler--Feynman action (\ref{Fey1}). The
mechanical--like locomotion term occurring at the single time $t$,
needs a covariant generalization from the flat $4-$dimensional
Euclidean space to the $n-$dimensional smooth Riemannian manifold,
so it becomes (see e.g., \cite{VladIJMMS,VladBrain},)
$$S[x]={1\over2}\int_{t_{ini}}^{t_{fin}}g_{ij}\,\dot{x}^{i}\dot{x}^{j}\,dt,\qquad \text{(summation
convention is always assumed)}
$$
where $g_{ij}$ is the Riemannian metric tensor that generates the
total `kinetic energy' of (loco)motions in the life space.

The second term in (\ref{Fey1}) gives the sophisticated definition
of Lewinian force--fields that drive the psychological
(loco)motions, if we interpret electrical charges $e_i$ occurring
at different times $t_i$ as motivational charges -- needs.

\subsubsection{Local Micro--Level of $LSF_{total}$}

After having properly defined macro--level MD \& CD, with a unique
transition map $F$ (including a unique motion path, driving field
and smooth--manifold geometry), we move down to the
\textit{microscopic} LSF--level of rapidly fluctuating MD \& CD,
where we cannot define a unique and smooth path--field--geometry.
The most we can do at this level of \textit{fluctuating
uncertainty}, is to formulate an adaptive path integral and
calculate overall probability amplitudes for ensembles of local
transitions from one LSF--point to the neighboring one. This
\textit{probabilistic transition micro--dynamics} is given by a
multi--path (field and geometry, respectively) and multi--phase
\textit{transition amplitude} $\langle Action|Intention\rangle$ of
corresponding to the globally--smooth transition map
(\ref{unique}). This absolute square of this probability amplitude
gives the \textit{transition probability} of occurring the final
state of $Action$ given the initial state of $Intention$,
$$P(Action|Intention)=|\langle Action|Intention\rangle|^2.
$$
The total transition amplitude from the state of $Intention$ to
the state of $Action$ is defined on $LSF_{total}$
\begin{equation}\langle Action|Intention\rangle_{total}:
INTENTION_{t_{0}}\cone{} ACTION_{t_{1}}\,,
\label{transdyn}\end{equation} given by adaptive generalization of
the Feynman's path integral
\cite{FeynmanOp,Feynman65,FeynmanSt,Feynman98}. The transition map
(\ref{transdyn}) calculates the \textit{overall probability
amplitude} along a multitude of wildly fluctuating paths, fields
and geometries, performing the \textit{microscopic} transition
from the micro--state $INTENTION_{t_{0}}$ occurring at initial
micro--time instant $t_{0}$ to the micro--state $ACTION_{t_{1}}$
at some later micro--time instant $t_{1}$, such that all
micro--time instants fit inside the global transition interval
$t_0,t_1,...,t_s\in[t_{ini},t_{fin}]$. It is symbolically written
as
\begin{equation}
\langle Action|Intention\rangle_{total} :=\put(0,0){\LARGE
$\int$}\put(20,15){\small $\Sigma$}\quad\,\mathcal{D}[w\Phi]\,
{\mathrm e}^{\mathrm i S[\Phi]}, \label{pathInt}
\end{equation}
where the Lebesgue integration is performed over all continuous
$\Phi^i_{con}=paths+fields+geometries$, while summation is
performed over all discrete processes and regional topologies
$\Phi^j_{dis}$. The symbolic differential $\mathcal{D}[w\Phi]$ in
the general path integral (\ref{pathInt}), represents an
\textit{adaptive path measure}, defined as a weighted product
\begin{equation}
\mathcal{D}[w\Phi]=\lim_{N\to\infty}\prod_{s=1}^{N}w_sd\Phi
_{s}^i, \qquad ({i=1,...,n=con+dis}),\label{prod}
\end{equation}
which is in practice satisfied with a large $N$ corresponding to
infinitesimal temporal division of the four motivational phases
(*).

Now, since Feynman's invention of the path integral
\cite{FeynmanOp}, a lot of research has been done to make the
real--time Feynman path integral mathematically rigorous (see
e.g., \cite{Loo,Loo2,Albeverio,Klauder,Klauder2,Klauder3}).

In the exponent of the path integral (\ref{pathInt}) we have the
action $S[\Phi]$ and the imaginary unit $\mathrm i=\sqrt{-1}$
($\mathrm i$ can be converted into the real number $-1$ using the
so--called \textit{Wick rotation}, see next subsection).

In this way, we get a range of micro--objects in the local LSF at
the short time--level: ensembles of rapidly fluctuating, noisy and
crossing paths, force--fields, local geometries with obstacles and
topologies with holes. However, by averaging process, both in time
and along ensembles of paths, fields and geometries, we can
recover the corresponding global MD and CD variables.

\subsubsection{Infinite--Dimensional Neural Network}

The adaptive path integral (\ref{pathInt}) incorporates the
\textit{local learning process} according to the basic formula
(see e.g., \cite{Grossberg,Grossberg2})
$$new\;value(t+1)\; =\; old\;value(t)\;+\; innovation(t)$$ The general \textit{weights}
$w_s=w_s(t)$ in (\ref{prod}) are updated by the $MONITOR$ feedback
during the transition process, according to one of the two
standard neural learning schemes, in which the micro--time level
is traversed in discrete steps, i.e., if $t=t_0,t_1,...,t_s$ then
$t+1=t_1,t_2,...,t_{s+1}$:
\begin{enumerate}
    \item A \textit{self--organized}, \textit{unsupervised}
    (e.g., Hebbian--like \cite{Hebb}) learning rule:
\begin{equation}
w_s(t+1)=w_s(t)+ \frac{\sigma}{\eta}(w_s^{d}(t)-w_s^{a}(t)),
\label{Hebb}
\end{equation}
where $\sigma=\sigma(t),\,\eta=\eta(t)$ denote \textit{signal} and
\textit{noise}, respectively, while superscripts $d$ and $a$
denote \textit{desired} and \textit{achieved} micro--states,
respectively; or
    \item A certain form of a \textit{supervised gradient descent
    learning}:
\begin{equation}
w_s(t+1)\,=\,w_s(t)-\eta \nabla J(t), \label{gradient}
\end{equation}
where $\eta $ is a small constant, called the \textit{step size},
or the \textit{learning rate,} and $\nabla J(n)$ denotes the
gradient of the `performance hyper--surface' at the $t-$th
iteration.
\end{enumerate}
Both Hebbian and supervised learning\footnote{Note that we could
also use a reward--based, \textit{reinforcement learning} rule
\cite{SB}, in which system learns its \textit{optimal policy}:
$$innovation(t)=|reward(t)-penalty(t)|.$$} are naturally used for the local decision making
process (see below) occurring at the intention formation faze
$\mathcal{F}$.

In this way, local micro--level of $LSF_{total}$ represents an
infinite--dimensional neural network. In the cognitive psychology
framework, our adaptive path integral (\ref{pathInt}) can be
interpreted as \textit{semantic integration} (see
\cite{Bransford,Ashcraft}).

\subsection{Pathways of (Loco)Motion and Decision Making in $LSF_{paths}$}

On the macro--level in the subspace $LSF_{paths}$ we have the
(loco)\textit{motion action principle}
$$
\delta S[x]=0,
$$
with the \textit{Newtonian--like action} $S[x]$ given by
\begin{equation}
S[x]=\int_{t_{ini}}^{t_{fin}}dt\,[{1\over2}g_{ij}\,\dot{x}^{i}\dot{x}^{j}+\varphi^i(x^i)],
\label{actmot}
\end{equation}
where $\dot{x}^i$ represents motivational (loco)motion velocity
vector with cognitive \textit{processing speed}. The first bracket
term in (\ref{actmot}) represents the kinetic energy $T$,
$$T={1\over2}g_{ij}\,\dot{x}^{i}\dot{x}^{j},$$ generated by the \textit{Riemannian
metric tensor} $g_{ij}$, while the second bracket term,
$\varphi^i(x^i)$, denotes the family of potential force--fields,
driving the (loco)motions $x^i=x^i(t)$ (the \textit{strengths} of
the fields $\varphi^i(x^i)$ depend on their positions $x^i$ in
LSF, see $LSF_{fields}$ below). The corresponding
Euler--Lagrangian equation gives the Newtonian--like equation of
motion
\begin{equation}
\frac{d}{dt}T_{\dot{x}^{i}}-T_{x^{i}}=-\varphi^i_{x^i},
\label{Newton}
\end{equation}
(subscripts denote the partial derivatives), which can be put into
the standard Lagrangian form
$$\frac{d}{dt}L_{\dot{x}^{i}}=L_{x^{i}},\qquad\text{with}\qquad
L=T-\varphi^i(x^i).
$$

Now, according to Lewin, the life space also has a sophisticated
topological structure. As a Riemannian smooth $n-$manifold, the
LSF--manifold $M$ gives rise to its fundamental
$n-$\emph{groupoid}, or $n-$category $\Pi _{n}(M)$ (see
\cite{Leinster,Leinster2}). In $\Pi _{n}(M)$, 0--cells are
\emph{points} in $M$; 1--cells are \emph{paths} in $M$ (i.e.,
parameterized smooth maps $f:[0,1]\rightarrow M$); 2--cells are \emph{%
smooth homotopies} (denoted by $\simeq $) \emph{of paths} relative
to endpoints (i.e., parameterized smooth maps $h:[0,1]\times
\lbrack 0,1]\rightarrow M$); 3--cells are \emph{smooth homotopies
of homotopies} of paths in $M$ (i.e., parameterized smooth maps
$j:[0,1]\times \lbrack 0,1]\times \lbrack 0,1]\rightarrow M$).
Categorical \emph{composition} is defined by \emph{pasting} paths
and homotopies. In this way, the following \textit{recursive
homotopy dynamics} emerges on the LSF--manifold $M$ (**):

\label{ivncat}
\begin{eqnarray*}
&&\mathtt{0-cell:}\,\,x_{0}\,\node\,\,\,\qquad x_{0}\in M; \qquad
\text{in
the higher cells below: }t,s\in[0,1]; \\
&&\mathtt{1-cell:}\,\,x_{0}\,\node\cone{f}\node\,x_{1}\qquad
f:x_{0}\simeq
x_{1}\in M, \\
&&f:[0,1]\rightarrow M,\,f:x_{0}\mapsto
x_{1},\,x_{1}=f(x_{0}),\,f(0)=x_{0},\,f(1)=x_{1}; \\
&&\text{e.g., linear path: }f(t)=(1-t)\,x_{0}+t\,x_{1};\qquad \text{or} \\
&&\text{Euler--Lagrangian }f-\text{dynamics with endpoint conditions }%
(x_0,x_1): \\
&&\frac{d}{dt}f_{\dot{x}^{i}}=f_{x^{i}},\quad \text{with}\quad
x(0)=x_{0},\quad x(1)=x_{1},\quad (i=1,...,n); \\
&&\mathtt{2-cell:}\,\,x_{0}\,\node\ctwodbl{f}{g}{h}\node\,x_{1}\qquad
h:f\simeq g\in M, \\
&&h:[0,1]\times \lbrack 0,1]\rightarrow M,\,h:f\mapsto g,\,g=h(f(x_{0})), \\
&&h(x_{0},0)=f(x_{0}),\,h(x_{0},1)=g(x_{0}),\,h(0,t)=x_{0},\,h(1,t)=x_{1} \\
&&\text{e.g., linear homotopy: }h(x_{0},t)=(1-t)\,f(x_{0})+t\,g(x_{0});\qquad%
\text{or} \\
&&\text{homotopy between two Euler--Lagrangian
}(f,g)-\text{dynamics}
\\
&&\text{with the same endpoint conditions }(x_0,x_1): \\
&&\frac{d}{dt}f_{\dot{x}^{i}}=f_{x^{i}},\quad \text{and} \quad \frac{d}{dt}%
g_{\dot{x}^{i}}=g_{x^{i}}\quad\text{with}\quad x(0)=x_{0},\quad
x(1)=x_{1};
\\
&&\mathtt{3-cell:}\,\,x_{0}\,\node\cthreecelltrp{f}{g}{h}{i}{j}\node%
\,x_{1}\qquad j:h\simeq i\in M, \\
&&j:[0,1]\times \lbrack 0,1]\times \lbrack 0,1]\rightarrow
M,\,j:h\mapsto
i,\,i=j(h(f(x_{0}))) \\
&&j(x_{0},t,0)=h(f(x_{0})),\,j(x_{0},t,1)=i(f(x_{0})), \\
&&j(x_{0},0,s)=f(x_{0}),\,j(x_{0},1,s)=g(x_{0}), \\
&&j(0,t,s)=x_{0},\,j(1,t,s)=x_{1} \\
&&\text{e.g., linear composite homotopy: }j(x_{0},t,s)=(1-t)\,h(f(x_{0}))+t%
\,i(f(x_{0})); \\
&&\text{or, homotopy between two homotopies between above two Euler-} \\
&&\text{Lagrangian }(f,g)-\text{dynamics with the same endpoint conditions }%
(x_0,x_1).
\end{eqnarray*}

In the next subsection we use the micro--level implications of
 the action $S[x]$ as given by (\ref{actmot}), for dynamical descriptions of the local
decision--making process.

On the micro--level in the subspace $LSF_{paths}$, instead of a
single path defined by the Newtonian--like equation of motion
(\ref{Newton}), we have an ensemble of fluctuating and crossing
paths with weighted probabilities (of the unit total sum). This
ensemble of micro--paths is defined by the simplest instance of
our adaptive path integral (\ref{pathInt}), similar to the
Feynman's original \textit{sum over histories},
\begin{equation}
\langle Action|Intention\rangle_{paths}=\put(0,0){\LARGE
$\int$}\put(20,15){\small $\Sigma$}\quad\, \mathcal{D}[wx]\,
\mathrm{e}^{{\mathrm i} S[x]},  \label{Feynman}
\end{equation}
where $\mathcal{D}[wx]$ is a functional measure on the
\textit{space of all weighted paths}, and the exponential depends
on the action $S[x]$ given by (\ref{actmot}). This procedure can
be redefined in a mathematically cleaner way if we Wick--rotate
the time variable $t$ to imaginary values, $t\mapsto \tau={\mathrm
i} t$, thereby making all integrals real:
\begin{equation}
\put(0,0){\LARGE $\int$}\put(20,15){\small $\Sigma$}\quad\,
\mathcal{D}[wx]\, \mathrm{e}^{{\mathrm i} S[x]}~\cone{Wick}\quad
\put(0,0){\LARGE $\int$}\put(20,15){\small $\Sigma$}\quad\,
\mathcal{D}[wx]\, \mathrm{e}^{-S[x]}. \label{Wick}
\end{equation}
Discretization of (\ref{Wick}) gives the standard
\textit{thermodynamic--like partition function}
\begin{equation}
Z=\sum_j{\mathrm e}^{-w_jE^j/T}, \label{partition}
\end{equation}
where $E^j$ is the motion energy eigenvalue (reflecting each
possible motivational energetic state), $T$ is the
temperature--like environmental control parameter, and the sum
runs over all motion energy eigenstates (labelled by the index
$j$). From (\ref{partition}), we can further calculate all
thermodynamic--like and statistical properties (see e.g., Feynman,
1972) of MD and CD, as for example, \textit{transition entropy},
$S = k_B\ln Z$, etc.

From cognitive perspective, our adaptive path integral
(\ref{Feynman}) calculates all (alternative) pathways of
information flow during the transition $Intention\to Action$.

In the language of transition--propagators, the integral over
histories (\ref{Feynman}) can be decomposed into the product of
propagators (i.e., Fredholm kernels or Green functions)
corresponding to the cascade of the four motivational phases (*)
\begin{equation}
\langle Action|Intention\rangle_{paths}=\put(0,0){\LARGE
$\int$}\put(20,15){\small $\Sigma$}\quad\,
dx^\mathcal{F}dx^\mathcal{I}dx^\mathcal{M}dx^\mathcal{T}
K(\mathcal{F},\mathcal{I})K(\mathcal{I},\mathcal{M})K(\mathcal{M},\mathcal{T}),
\label{prop}
\end{equation}
satisfying the Schr\"{o}dinger--like equation (see e.g.,
\cite{Dirac})
\begin{equation}
{\mathrm i}\,\partial_t\langle
Action|Intention\rangle_{paths}=H_{Action}\,\langle
Action|Intention\rangle_{paths}, \label{Schr}
\end{equation}
where $H_{Action}$ represents the Hamiltonian (total energy)
function available at the state of $Action$. Here our `golden
rule' is: the higher the Hamiltonian $H_{Action}$, the lower the
microscopic fatigue.

In the connectionist language, our propagator expressions
(\ref{prop}--\ref{Schr}) represent \textit{activation dynamics},
to which our $Monitor$ process gives a kind of
\textit{backpropagation} feedback, a common type of supervised
learning (\ref{gradient}).

\subsubsection{Mechanisms of decision making under uncertainty}

Now, the basic question about our local decision making process,
occurring under uncertainty at the intention formation faze
$\mathcal{F}$, is: Which alternative to choose? (see
\cite{Roe,Grossberg2,Ashcraft}). In our path--integral language
this reads: Which path (alternative) should be given the highest
probability weight $w$? Naturally, this problem is iteratively
solved by the learning process (\ref{Hebb}--\ref{gradient}),
controlled by the $MONITOR$ feedback, which we term
\textit{algorithmic approach}.

In addition, here we analyze qualitative mechanics of the local
decision making process under uncertainty, as a \textit{heuristic
approach}. This qualitative analysis is based on the micro--level
interpretation of the Newtonian--like action $S[x]$, given by
(\ref{actmot}) and figuring both processing speed $\dot{x}$ and
LTM (i.e., the force--field $\varphi(x)$, see next subsection).
Here we consider three different cases:\begin{enumerate}
    \item If the potential $\varphi(x)$ is not very dependent upon
    position $x(t)$, then the more direct paths contribute the
    most, as longer paths, with higher mean square velocities
    $[\dot{x}(t)]^2$ make the exponent more negative (after Wick rotation
    (\ref{Wick})).

    \item On the other hand, suppose that $\varphi(x)$ does indeed
    depend on position $x$. For simplicity, let the potential
    increase for the larger values of $x$. Then a direct path does
    not necessarily give the largest contribution to the overall
    transition probability, because the integrated value of the
    potential is higher than over another paths.

    \item Finally, consider a path that deviates widely from the
    direct path. Then $\varphi(x)$ decreases over that path, but at the
    same time the velocity $\dot{x}$ increases. In this case, we
    expect that the increased velocity $\dot{x}$ would more than
    compensate for the decreased potential over the path.
\end{enumerate}
Therefore, the most important path (i.e., the path with the
highest weight $w$) would be one for which any smaller integrated
value of the surrounding field potential $\varphi(x)$ is more than
compensated for by an increase in kinetic--like energy
${m\over2}\dot{x}^2$. In principle, this is neither the most
direct path, nor the longest path, but rather a middle way between
the two. Formally, it is the path along which the average
Lagrangian is minimal,
\begin{equation}
<{m\over2}\dot{x}^2+\varphi(x)>~\longrightarrow~\min, \label{DM}
\end{equation}
i.e., the \textit{path that requires minimal memory} (both LTM and
WM, see $LSF_{fields}$ below) and \textit{processing speed}. This
mechanical result is consistent with the `filter theory' of
\textit{selective attention} \cite{Broadbent}, proposed in an
attempt to explain a range of the existing experimental results.
This theory postulates a low level filter that allows only a
limited number of percepts to reach the brain at any time. In this
theory, the importance of conscious, directed attention is
minimized. The type of attention involving low level filtering
corresponds to the concept of \textit{early selection}
\cite{Broadbent}.

Although we termed this `heuristic approach' in the sense that we
can instantly feel both the processing speed $\dot{x}$ and the LTM
field $\varphi(x)$ involved, there is clearly a psycho--physical
rule in the background, namely the averaging minimum relation
(\ref{DM}).

From the decision making point of view, all possible paths
(alternatives) represent the \textit{consequences} of decision
making. They are, by default, \textit{short--term consequences},
as they are modelled in the micro--time--level. However, the path
integral formalism allows calculation of the \textit{long--term
consequences}, just by extending the integration time,
$t_{fin}\to\infty$. Besides, this \textit{averaging decision
mechanics} -- choosing the optimal path -- actually performs the
`averaging lift' in the LSF: from micro--level to the
macro--level.

\subsection{Force--Fields and Memory in $LSF_{fields}$}

At the macro--level in the subspace $LSF_{fields}$ we formulate
the \textit{force--field action principle}
\begin{equation}
\delta S[\varphi]=0, \label{acprf}
\end{equation}
with the action $S[\varphi]$ dependent on Lewinian force--fields
$\varphi^{i}=\varphi^{i}(x)\,\,(i=1,...,N)$, defined as a temporal
integral
\begin{equation}
S[\varphi]=\int_{t_{ini}}^{t_{fin}}\mathfrak{L}[\varphi]\,dt,
\label{actf}
\end{equation}
with Lagrangian density given by
\begin{equation*}
\mathfrak{L}[\varphi]=\int
d^{n}x\,\mathcal{L}(\varphi_i,\partial_{x^j}\varphi^i),
\end{equation*}%
where the integral is taken over all $n$ coordinates $x^j=x^j(t)$
of the LSF, and $\partial_{x^j}\varphi^i$ are partial derivatives
of the field variables over coordinates.

On the micro--level in the subspace $LSF_{fields}$ we have the
Feynman--type \textit{sum over fields}
$\varphi^{i}\,\,(i=1,...,N)$ given by the adaptive path integral
\begin{equation}
\langle Action|Intention\rangle_{fields} =\put(0,0){\LARGE
$\int$}\put(20,15){\small $\Sigma$}\quad\,\mathcal{D}[w\varphi]\,
{\mathrm e}^{\mathrm i} S[\varphi]~ \cone{Wick}\quad
\put(0,0){\LARGE $\int$}\put(20,15){\small
$\Sigma$}\quad\,\mathcal{D}[w\varphi]\, {\mathrm e}^{-
S[\varphi]}, \label{field}
\end{equation}
with action $S[\varphi]$ given by temporal integral (\ref{actf}).
(Choosing special forms of the force--field action $S[\varphi]$ in
(\ref{field}) defines micro--level MD \& CD, in the LSF$_{fields}$
space, that is similar to standard quantum--field equations, see
e.g., \cite{Ramond}.) The corresponding partition function has the
form similar to (\ref{partition}), but with field energy levels.

Regarding topology of the force fields, we have in place
$n-$\emph{categorical} \textit{Lagrangian--field structure} on the
Riemannian LSF manifold $M$,
$$\Phi^{i}:[0,1]\rightarrow M,\,\Phi^{i}:\Phi^{i}_0\mapsto
\Phi^{i}_1,$$ generalized from (**) above, using
$$\frac{d}{dt}f_{\dot{x}^{i}}=f_{x^{i}}
\longrightarrow\partial _{\mu }\left( \frac{\partial
\mathcal{L}}{\partial _{\mu }\Phi^{i}}\right)=\frac{\partial
\mathcal{L}}{\partial \Phi^{i}},$$ with
$$[x_0,x_1]\longrightarrow[\Phi^{i}_0,\Phi^{i}_1].
$$

\subsubsection{Relationship between memory and force--fields}

As already mentioned, the subspace $LSF_{fields}$ is related to
our \textit{memory storage} \cite{Ashcraft}. Its global
macro--level represents the \textit{long--term memory} (LTM),
defined by the least action principle (\ref{acprf}), related to
\textit{cognitive economy} in the model of \textit{semantic
memory}  \cite{Collins}. Its local micro--level represents
\textit{working memory} (WM), a limited--capacity `bottleneck'
defined by the adaptive path integral (\ref{field}). According to
our formalism, each of Miller's $7\pm 2$ units \cite{Miller} of
the local WM are adaptively stored and averaged to give the global
LTM capacity (similar to the physical notion of potential). This
averaging memory lift, from WM to LTM represents
\textit{retroactive interference}, while the opposite direction,
given by the path integral (\ref{field}) itself, represents
\textit{proactive interference}. Both retroactive and proactive
interferences are examples of the impact of cognitive contexts on
memory. Motivational contexts can exert their influence, too. For
instance, a reduction in task--related recall following the
completion of the task -- the well--known \textit{Zeigarnik
effect} \cite{Zeigarnik} -- is one of the clearest examples of
force--field influences on memory: the amount of details
remembered of a task declines as the force--field tension to
complete the task is reduced by actually completing it.

Once defined, the global LTM potential $\varphi=\varphi(x)$ is
then affecting the locomotion transition paths through the path
action principle (\ref{actmot}), as well as general learning
(\ref{Hebb}--\ref{gradient}) and decision making process
(\ref{DM}).

On the other hand, the two levels of $LSF_{fields}$ fit nicely
into the two levels of processing framework, as presented by
\cite{Craik}, as an alternative to theories of separate stages for
sensory, working and long--term memory. According to the
\textit{levels of processing framework}, stimulus information is
processed at multiple levels simultaneously depending upon its
characteristics. In this framework, our macro--level memory field,
defined by the fields action principle (\ref{acprf}), corresponds
to the \textit{shallow memory}, while our micro--level memory
field, defined by the adaptive path integral (\ref{field}),
corresponds to the \textit{deep memory}.

\subsection{Geometries, Topologies and Noise in $LSF_{geom}$}

On the macro--level in the subspace $LSF_{geom}$ representing an
$n-$dimensional smooth manifold $M$ with the global Riemannian
metric tensor $g_{ij}$, we formulate the \textit{geometric action
principle}
$$
\delta S[g_{ij}]=0,
$$
where $S=S[g_{ij}]$ is the $n-$dimensional \textit{geodesic
action} on $M$,
\begin{equation}
S[g_{ij}]=\int d^n x\sqrt{g_{ij}\,dx^{i}dx^{j}}. \label{geod}
\end{equation}
The corresponding Euler--Lagrangian equation gives the
\textit{geodesic equation} of the \textit{shortest path} in the
manifold $M$,
\[
\ddot{x}^{i}+\Gamma _{jk}^{i}\,\dot{x}^{j}\,\dot{x}^{k}=0,
\]
where $\Gamma _{jk}^{i}$ are the Christoffel's symbols of the
affine connection on $M$, which is the source of the
\textit{curvature} of $M$, and at the same time is a geometrical
description for \textit{noise} (see \cite{Ingber,Ingber2}). The
higher the local curvatures of the LSF--manifold $M$, the greater
the noise in the life space. This noise is the source of our
micro--level fluctuations. It can be internal or external; in both
cases it curves our micro--LSF.

Otherwise, if instead we choose an $n-$dimensional Hilbert--like
action (see \cite{Misner}),
\begin{equation}
S[g_{ij}]=\int d^n x\sqrt{\det|g_{ij}}|R, \label{Hilbert}
\end{equation}
where $R$ is the scalar curvature (derived from $\Gamma
_{jk}^{i}$), we get the $n-$dimensional Einstein-like equation
$$
G_{ij}=8\pi T_{ij},
$$
where $G_{ij}$ is the Einstein--like tensor representing geometry
of the LSF manifold $M$ ($G_{ij}$ is the trace--reversed Ricci
tensor $R_{ij}$, which is itself the trace of the Riemann
curvature tensor of the manifold $M$), while $T_{ij}$ is the
$n-$dimensional \textit{stress--energy--momentum} tensor. This
equation explicitly states that \textit{psycho--physics of the LSF
is proportional to its geometry}. $T_{ij}$ is important quantity,
representing motivational \textit{energy}, geometry--imposed
\textit{stress} and \textit{momentum} of (loco)motion. As before,
we have our `golden rule': \textit{the greater the}
$T_{ij}-$\textit{components, the higher the speed of cognitive
processes and the lower the macroscopic fatigue.}

The choice between the geodesic action (\ref{geod}) and the
Hilbert action (\ref{Hilbert}) depends on our interpretation of
time. If time is not included in the LSF manifold $M$
(non--relativistic approach) then we choose the geodesic action.
If time is included in the LSF manifold $M$ (making it a
relativistic--like $n-$dimensional space--time) then the Hilbert
action is preferred. The first approach is more related to the
information processing and the working memory. The later,
space--time approach can be related to the long--term memory: we
usually recall events closely associated with the times of their
happening.

On the micro--level in the subspace $LSF_{geom}$ we have the
adaptive \textit{sum over geometries}, represented by the path
integral over all local (regional) Riemannian metrics
$g_{ij}=g_{ij}(x)$ varying from point to point on $M$ (modulo
diffeomorphisms),
\begin{equation}
\langle Action|Intention\rangle_{geom}=\put(0,0){\LARGE
$\int$}\put(20,15){\small
$\Sigma$}\quad\, \mathcal{D}[wg_{ij}]\,\mathrm{e}%
^{{\mathrm i} S[g_{ij}]}~ \cone{Wick}\quad \put(0,0){\LARGE
$\int$}\put(20,15){\small $\Sigma$}\quad\,\mathcal{D}[wg_{ij}]\,
{\mathrm e}^{- S[g_{ij}]}, \label{geom}
\end{equation}
where $\mathcal{D}[g_{ij}]$ denotes diffeomorphism equivalence
classes of metrics $g_{ij}(x)$ of $M$.

To include the topological structure (e.g., a number of holes) in
$M$, we can extend (\ref{geom}) as\begin{equation} \langle
Action|Intention\rangle_{geom/top}=\sum_{\mathrm{topol.}}\put(0,0){\LARGE
$\int$}\put(20,15){\small $\Sigma$}\quad\,
\mathcal{D}[wg_{ij}]\,\mathrm{e}^{{\mathrm i} S[g_{ij}]},
\label{top}
\end{equation}
where the topological sum is taken over all
connectedness--components of $M$ determined by the \textit{Euler
characteristics} of $M$. This type of integral defines the
\textit{theory of fluctuating geometries}, a propagator between
$(n-1)-$dimensional boundaries of the $n-$dimensional manifold
$M$. One has to contribute a meaning to the integration over
geometries. A key ingredient in doing so is to approximate (using
simplicial approximation and Regge calculus \cite{Misner}) in a
natural way the smooth structures of the manifold $M$ by piecewise
linear structures (mostly using topological simplices $\Delta$).
In this way, after the Wick--rotation (\ref{Wick}), the integral
(\ref{geom}--\ref{top}) becomes a \textit{simple statistical
system}, given by partition function
\begin{equation*}
Z=\sum_{\Delta}\frac{1}{C_{\Delta}}e^{-S_{\Delta}},
\end{equation*}%
where the summation is over all triangulations $\Delta$ of the
manifold $M$, while the number $C_{T}$ is the order of the
automorphism group of the performed triangulation.

\subsubsection{Micro--level geometry: the source of noise and stress in LSF}

The subspace $LSF_{geom}$ is the source of noise, fluctuations and
obstacles, as well as psycho--physical stress. Its micro--level is
adaptive, reflecting the human ability to efficiently act within
the noisy environment and under the stress conditions. By
averaging it produces smooth geometry of certain curvature, which
is at the same time the smooth psycho--physics. This macro--level
geometry directly affects the memory fields and indirectly affects
the (loco)motion transition paths.

\section{Discussion}

We have presented a new psychodynamical concept of the life space
foam, as a natural medium for both motivational dynamics and
induced cognitive theory of learning, memory, information
processing and decision making. Its macro--level has been defined
using the least action principle, while its micro--level has been
defined using adaptive path integral. The totality of six facets
of the LSF have been presented: paths, fields and geometries, on
both levels.

The formalisms proposed and developed in this paper, can be
employed in generating a number of meaningful and  useful
predictions about motivational and cognitive dynamics. These
predictions range from effects of fatigue or satiation on
goal-directed performance, through general learning and memory
issues, to the sophisticated selection between conflicting
alternatives at decision making and/or sustained action stages.

For example, one of the simplest types of performance--degrading
disturbances in the LSF is what we term motivational fatigue -- a
motivational drag factor that slows the actors' progress towards
their goal. There are two fundamentally different sources of this
motivational drag, both leading to apparently the same reduction
in performance:  (a) tiredness / exhaustion and (b) satiation
(e.g., boredom). Both involve the same underlying mechanism (the
raising valence of the alternatives to continuing the action) but
the alternatives will differ considerably, depending on the
properties of the task, from self--preservation / recuperation in
the exhaustion case through to competing goals in the satiation
case.

The spatial representation of this motivational drag is relatively
simple: uni--dimensional LSF--coordinates may be sufficient for
most purposes, which makes it attractive for the initial
validation of our predictive model. Similarly uncomplicated
spatial representations can be achieved for what we term
motivational boost derived from the proximity to the goal
(including the well--known phenomenon of `the home stretch'): the
closer the goal (e.g., a finishing line) is perceived to be, the
stronger its `pulling power' \cite{Lewin51,Lewin97,Aidman}.
Combinations of motivational drag and motivational boost effects
may be of particular interest in a range of applications. These
combinations can be modelled within relatively simple
uni--dimensional LSF--coordinate systems.

In their general form, the formalisms developed in this paper, are
consistent with dynamic--connectionist models of decision making,
such as decision field theory \cite{Busemeyer,Busemeyer2} and
related theory of memory retrieval \cite{Ratcliff}. In particular,
our multi--path integrals are able, in principle, to account for
multi--alternative preferential choice behaviors as conceptualized
in the multi--alternative decision field theory \cite{Roe},
potentially leading to testable predictions, e.g., those
concerning the effects of similarity, compromise and time pressure
on decision quality.

Examining specific predictions of this type in various
goal--oriented contexts is the focus of our current work.

\end{document}